\documentclass[pdflatex,sn-nature,referee,a4paper]{sn-jnl}% Style 

\usepackage{graphicx}%
\usepackage{multirow}%
\usepackage{amsmath,amssymb,amsfonts}%
\usepackage{amsthm}%
\usepackage{mathrsfs}%
\usepackage[title]{appendix}%
\usepackage{xcolor}%
\usepackage{textcomp}%
\usepackage{manyfoot}%
\usepackage{booktabs}%
\usepackage{algorithm}%
\usepackage{algorithmicx}%
\usepackage{algpseudocode}%
\usepackage{listings}%
\usepackage{anyfontsize}%
\usepackage[separate-uncertainty=true,multi-part-units=single]{siunitx}%
\usepackage{physics}%

\newcommand{\Vcut}{V_\mathrm{cut}}
\newcommand{\VQDL}{V_\mathrm{LD}}
\newcommand{\VQDR}{V_\mathrm{RD}}
\newcommand{\VCS}{V_\mathrm{CS}}
\newcommand{\VH}{V_\mathrm{H}}
\newcommand{\muQDL}{\mu_\mathrm{LD}}
\newcommand{\muQDR}{\mu_\mathrm{RD}}

\newcommand{\VMRF}{V_\mathrm{M}^\mathrm{RF}}
\newcommand{\VRRF}{V_\mathrm{R}^\mathrm{RF}}
\newcommand{\VLDC}{V_\mathrm{L}^\mathrm{DC}}
\newcommand{\VRDC}{V_\mathrm{R}^\mathrm{DC}}
\newcommand{\SM}{S_\mathrm{M}^\mathrm{11}}
\newcommand{\SR}{S_\mathrm{R}^\mathrm{11}}

\begin{document}

\title[Article Title]{Single-shot parity readout of a minimal Kitaev chain}

\author*[1]{\fnm{Nick} \spfx{van} \sur{Loo}}
\email{n.vanloo@tudelft.nl}
\equalcont{These authors contributed equally to this work.}

\author[1]{\fnm{Francesco} \sur{Zatelli}}
\equalcont{These authors contributed equally to this work.}

\author[2]{\fnm{Gorm O.} \sur{Steffensen}}
\equalcont{These authors contributed equally to this work.}

\author[1]{\fnm{Bart} \sur{Roovers}}
\equalcont{These authors contributed equally to this work.}

\author[1]{\fnm{Guanzhong} \sur{Wang}}
\author[1]{\fnm{Thomas} \spfx{Van} \sur{Caekenberghe}}
\author[1]{\fnm{Alberto} \sur{Bordin}}
\author[1]{\fnm{David} \spfx{van} \sur{Driel}}
\author[1]{\fnm{Yining} \sur{Zhang}}
\author[1]{\fnm{Wietze D.} \sur{Huisman}}
\author[3]{\fnm{Ghada} \sur{Badawy}}

\author[3]{\fnm{Erik P.A.M.} \sur{Bakkers}}
\author[1]{\fnm{Grzegorz P.} \sur{Mazur}}
\author[2]{\fnm{Ramón} \sur{Aguado}}
\author*[1]{\fnm{Leo P.} \sur{Kouwenhoven}}
\email{l.p.kouwenhoven@tudelft.nl}

\affil[1]{\orgdiv{QuTech and Kavli Institute of Nanoscience}, \orgname{Delft University of Technology},

\country{The Netherlands}}

\affil[2]{\orgdiv{Instituto de Ciencia de Materiales de Madrid (ICMM)},

\orgname{Consejo Superior de Investigaciones Científicas (CSIC)}, \country{Spain}}

\affil[3]{\orgdiv{Department of Applied Physics}, \orgname{Eindhoven University of Technology},

\country{The Netherlands}}

%%==================================%%
%% Abstract %%
%%==================================%%

\abstract{Protecting qubits from noise is essential for building reliable quantum computers. Topological qubits offer a route to this goal by encoding quantum information non-locally, using pairs of Majorana zero modes. These modes form a shared fermionic state whose occupation -- either even or odd -- defines the fermionic parity that encodes the qubit~\cite{kitaev2001unpaired}. Importantly, this parity can only be accessed by a measurement that couples two Majoranas to each other. A promising platform for realizing such qubits is the Kitaev chain~\cite{kitaev2001unpaired}, implemented in quantum dots coupled via superconductors~\cite{sau2012realizing}. Even the minimal two-site chain hosts a pair of Majorana modes, often called poor man’s Majoranas, which are spatially separated but offer limited protection compared to longer chains~\cite{leijnse2012parity, dvir2023pmm, tenhaaf2024pmm}. Here we introduce a measurement technique that reads out their parity through quantum capacitance. Our method couples two Majoranas and resolves their parity in real time, visible as random telegraph switching with lifetimes exceeding a millisecond. Simultaneous charge sensing confirms that the two parity states are charge neutral and remain indistinguishable to a probe that does not couple the modes. These results establish the essential readout step for time-domain control of Majorana qubits, resolving a long-standing experimental challenge.

\par\bigskip\bigskip\bigskip\bigskip\bigskip\bigskip\bigskip\bigskip\bigskip\bigskip\bigskip\bigskip\bigskip\bigskip\bigskip\bigskip\bigskip\bigskip\bigskip
{\centering\footnotesize
Published in \textit{Nature} \textbf{650}, 334--339 (2026).
\href{https://doi.org/10.1038/s41586-025-09927-7}
{doi:10.1038/s41586-025-09927-7}.
The definitive version is available from Nature.
\par}}

\maketitle
\pagebreak

\section*{Introduction}\label{Intro}

In the quest for fault-tolerant quantum computing, Majorana zero modes offer an attractive platform for encoding topological qubits. These qubits, defined by the fermionic parity of spatially separated pairs of Majorana modes, exhibit intrinsic protection against decoherence~\cite{kitaev2001unpaired,kitaev2003ftqc,nayak2008topoqc,sarma2015topoqc}. The simplest model supporting these modes is the Kitaev chain: a one-dimensional array of spinless fermions with superconducting pairing~\cite{kitaev2001unpaired}. Their recent realization in quantum dots (QDs) coupled via superconductors has opened an exciting new frontier~\cite{dvir2023pmm,tenhaaf2024pmm}. Even the minimal version of these chains hosts a pair of Majorana zero modes at a fine-tuned sweet spot that are often referred to as ``poor man’s Majoranas" (PMMs)~\cite{sau2012realizing,leijnse2012parity}. At that sweet spot, the even and odd parity states of the chain are degenerate, supporting Majorana modes $\gamma_1$ and $\gamma_2$ that localize on different quantum dots. Their spatial separation provides protection against detuning of either dot. However, this protection is fragile: the parity degeneracy splits linearly when perturbing the coupling between the dots, while simultaneous detuning of both QDs lifts it quadratically~\cite{leijnse2012parity}. In longer chains, by contrast, the residual splitting from noise or imperfect tuning decreases exponentially with chain length, which is the defining feature of topological protection~\cite{sau2012realizing}. Despite their limited robustness, PMMs are expected to exhibit non-Abelian exchange statistics~\cite{liu2023fusion,boross2024braiding,tsintzis2024majorana,seoane2024subgap}.

\begin{figure*}[h!]
    \centering
    \includegraphics[width=1\textwidth]{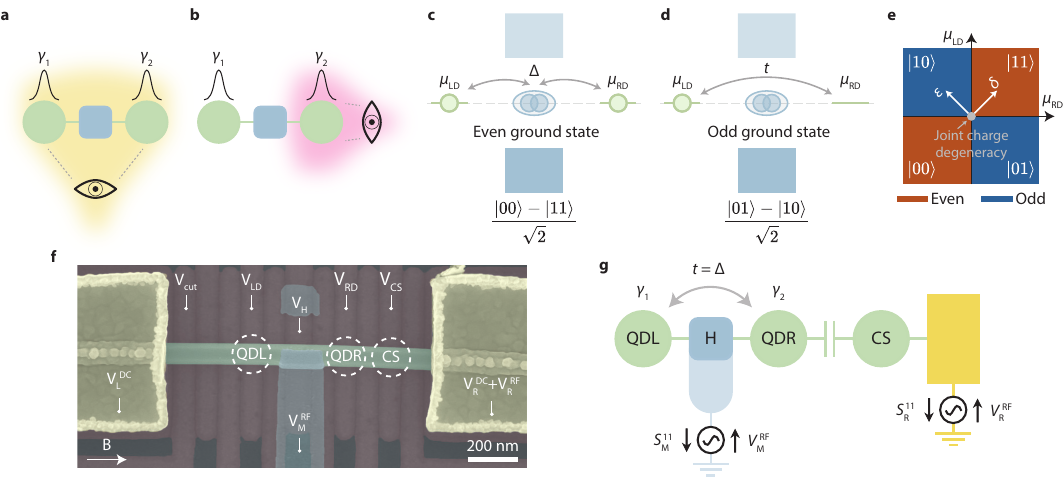}
    \caption{\textbf{Minimal Kitaev chain and parity measurement setup.}
    \textbf{a,} Illustration of two Majorana modes $\gamma_1$ and $\gamma_2$ on separate dots. A probe acting on both couples them and reveals their parity.
    \textbf{b,} Illustration of a probe acting on only one Majorana without inducing a coupling, so that parity remains indistinguishable.
    \textbf{c,} Energy diagram of two quantum dots with chemical potentials $\muQDL$ and $\muQDR$ in the even parity sector, where $|00\rangle$ and $|11\rangle$ hybridize via crossed Andreev reflection (CAR) in the superconductor with amplitude $\Delta$. 
    \textbf{d,} Energy diagram of the odd parity sector, where $|10\rangle$ and $|01\rangle$ hybridize via elastic co-tunneling (ECT) with amplitude $t$.
    \textbf{e,} Charge-stability diagram of two quantum dots, showing the detuning axis $\varepsilon$ and common-mode axis $\delta$. Orange and blue sectors denote the even and odd parity ground states. The center marks the joint charge degeneracy, which is the PMM sweet spot when $t=\Delta$.
    \textbf{f,} False-colored scanning electron micrograph of the device. The InSb nanowire (green) is proximitized by an Al strip (blue) and contacted by Cr/Au leads (yellow). Quantum dots QDL, QDR and CS (dashed circles) are defined using Ti/Pd bottom gates (red). The Kitaev chain is tuned by $\VQDL$, $\VQDR$ and $\VH$. The device is isolated from the left lead using $\Vcut$. The charge sensor (CS) is controlled with $\VCS$ and the nanowire is depleted between QDR and CS. DC ($\VLDC$ and $\VRDC$) and RF ($\VMRF$ and $\VRRF$) voltages are applied to the three leads for biasing and reflectometry. A parallel magnetic field $B = 150\,\mathrm{mT}$ ensures spin polarization of the QDs.  
    \textbf{g,} Schematic of the readout setup. The two QDs are coupled through a superconducting segment connected to an RF resonator, for readout via $\SM$. Likewise, the CS is read out via a second resonator $\SR$.
    }
    \label{fig:1}
\end{figure*}

In general, two Majorana modes form a fermion $c = (\gamma_1 + i\gamma_2)/\sqrt{2}$ whose occupation $P_{12} = i\gamma_1\gamma_2$ stores the quantum information~\cite{bonderson2008measurement,vijay2016teleportation,plugge2017majorana,tsintzis2024majorana}. Accessing this parity requires coupling the modes to an observable of a detector $M_\mathrm{detector}$ (e.g., the photonic mode of a resonator), introducing an interaction of the form $i\gamma_1\gamma_2 M_\mathrm{detector}$ as schematically depicted in Fig.~\ref{fig:1}a,b. This controlled coupling hybridizes the modes and fuses them into an ordinary fermionic level~\cite{sau2025capacitance}. Parity readout then reduces to measuring the occupation of a fermion in a semiconductor–superconductor hybrid system. Such measurements have been extensively explored using charge sensing~\cite{menard2019suppressing}, dispersive gate sensing~\cite{sabonis2019dispersive,deJong2023controllable} and circuit quantum electrodynamics~\cite{hays2018direct,bargerbos2022singlet,hinderling2024flip}. Recently, single-shot readout was demonstrated in an interferometry-based implementation exhibiting $h/2e$-periodic bimodality and millisecond-scale lifetimes~\cite{aghaee2025interferometric}.
The same approach was employed in a multi-wire geometry to extract loop-specific parity-switching times~\cite{aghaee2025distinct}.

Beyond readout, a central challenge is to preserve parity over time. In practice, conservation of parity can be broken by quasiparticles entering the device, a process known as quasiparticle poisoning. This causes a Majorana qubit to leak out of the computational subspace, destroying the stored quantum information~\cite{rainis2012majorana}. Similar poisoning limits the performance of superconducting and hybrid qubits~\cite{aumentado2004nonequilibrium,catelani2011relaxation,hays2018direct}. Characterizing parity lifetimes and achieving high-fidelity parity readout therefore remain critical benchmarks toward Majorana-based qubits and topological quantum computing~\cite{rainis2012majorana,karzig2021quasiparticle,aghaee2025interferometric}.

In this work, we introduce a measurement technique that distinguishes the fermionic parity of a minimal Kitaev chain through its quantum capacitance, by detuning the poor man’s Majoranas around the sweet spot and thereby inducing a parity-dependent splitting~\cite{contamin2021hybrid,liu2023fusion}. Unlike earlier dispersive sensing of charge states in quantum dots and hybrid systems~\cite{colless2013dispersive,deJong2019rapid,vigneau2023probing}, our method detects the response directly through the superconducting lead. This reflects the common-mode motion of charge between the dots and the superconductor, providing a new route to dispersive readout that resolves the parity of the chain in real time. By comparing with simultaneous charge sensing of one QD, we further demonstrate that the detected parity states are charge neutral.

\section*{Minimal Kitaev chain quantum capacitance}\label{TransportPMM}
A minimal Kitaev chain consists of two spinless fermionic sites with on-site energies $\muQDL$ and $\muQDR$. The sites are coupled by elastic co-tunneling (ECT) and crossed Andreev reflection (CAR), with amplitudes $t$ and $\Delta$, respectively. Each site can be either occupied or empty, giving rise to four charge configurations $|n_\mathrm{LD} n_\mathrm{RD}\rangle$, where $n_\mathrm{LD},n_\mathrm{RD} \in \{0,1\}$. At charge degeneracy, the even states $|00\rangle$ and $|11\rangle$ hybridize via CAR to form bonding and anti-bonding combinations $|e\rangle = (|00\rangle \pm |11\rangle)/\sqrt{2}$ (Fig.~\ref{fig:1}c). Likewise, the odd  states $|10\rangle$ and $|01\rangle$ hybridize via ECT to form $|o\rangle = (|10\rangle \pm |01\rangle)/\sqrt{2}$ (Fig.~\ref{fig:1}d). These define two parity branches with distinct energy dispersions:
\begin{align}
E_\mathrm{e}^\pm &= -\delta \pm \sqrt{\delta^2 + \Delta^2}, \label{eq:Ee} \\
E_\mathrm{o}^\pm &= -\delta \pm \sqrt{\varepsilon^2 + t^2}. \label{eq:Eo}
\end{align}
Here, $E_\mathrm{e}^\pm$ and $E_\mathrm{o}^\pm$ are the even and odd eigenenergies with $\pm$ denoting ground and excited states, $\varepsilon = (\muQDL-\muQDR)/2$ defines the detuning axis and $\delta = (\muQDL+\muQDR)/2$  the common-mode axis. When $t = \Delta$ and $\muQDL = \muQDR = 0$, the two ground states become degenerate, giving rise to zero-energy excitations with Majorana character on each dot~\cite{leijnse2012parity}. The corresponding charge-stability diagram is shown in Fig.~\ref{fig:1}e, where the PMM sweet spot is located at the joint charge degeneracy.

Our realization (Fig.~\ref{fig:1}f) consists of two gate-defined quantum dots (QDL and QDR) that form a minimal Kitaev chain in an InSb–Al hybrid nanowire~\cite{badawy2019high}, with a third dot acting as a nearby charge sensor (CS)~\cite{persson2010excess,bruhat2016cavity,vandrielPRXQuantum.5.020301}. While transport measurements are typically used to set up the Kitaev chain~\cite{dvir2023pmm}, maintaining the required coupling to normal leads during readout would continuously inject quasiparticles that flip the fermion parity. To preserve parity, we instead isolate the device by depleting the nanowire with $V_\mathrm{cut}$. The isolated system is then probed through the reflected signal of two multiplexed RF resonators (Fig.~\ref{fig:1}g): $\SM$ from the superconducting lead, sensitive to the parity of the chain through quantum capacitance, and $\SR$ from the nearby charge sensor that monitors local charge. We analyze these signals projected along their principal axes as $\Delta \SR$ and $\Delta \SM$. Full fabrication and wiring details are given in Methods.

\begin{figure*}[h!]
    \centering
    \includegraphics[width=0.5\textwidth]{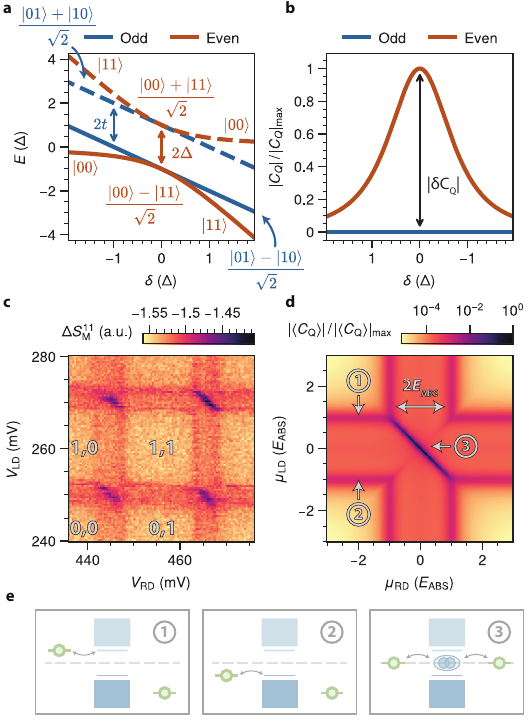}
    \caption{\textbf{Quantum capacitance of the chain.}
    \textbf{a,} Energy dispersion of the even (CAR-coupled) and odd (ECT-coupled) parity states at zero detuning ($\varepsilon=0$), plotted as a function of the common-mode potential $\delta$ along which $\VMRF$ is applied.
    \textbf{b,} Calculated quantum capacitance $C_\mathrm{Q}$ for each parity branch, given by the second derivative of energy with respect to chemical potential. A strong contrast $\delta C_\mathrm{Q}$ appears near the degeneracy point where even states show maximal curvature.
    \textbf{c,} Charge-stability diagram with labeled charge states measured via the quantum capacitance signal $\Delta \SM$. The dominant anti-diagonal lines correspond to $\muQDL=-\muQDR$, where the strength of CAR is maximal.  
    \textbf{d,} Simulation of the quantum capacitance around one charge degeneracy (see Methods), showing features consistent with the experiment.
    \textbf{e,} Energy diagrams of the relevant hybridization processes. (1) and (2) correspond to hybridization between a single dot and the lowest-energy state in the hybrid segment; (3) shows interdot hybridization at joint charge degeneracy. 
}
    \label{fig:2}
\end{figure*}

We first focus on the superconducting lead. The applied RF signal induces small oscillations in its electrochemical potential relative to the QDs, effectively modulating the system along the common-mode axis $\delta$. Such oscillations could also weakly shift the energies of Andreev bound states and thereby affect the effective couplings, but this effect is negligible due to strong capacitive coupling between the ABSs and the superconductor. As seen from Eqs.~\ref{eq:Ee} and \ref{eq:Eo}, this reveals a key difference between the parity branches: even states disperse nonlinearly due to hybridization via CAR, while the odd states change linearly (Fig.~\ref{fig:2}a). This is reflected in the curvature of their energy levels and thus in their quantum capacitance, which is proportional to the second derivative of energy with respect to chemical potential (see Methods). Intuitively, the RF oscillation periodically shifts the balance between $|00\rangle$ and $|11\rangle$, inducing charge motion between the quantum dots and the superconductor and producing a measurable quantum capacitance. In contrast, the odd states $|01\rangle$ and $|10\rangle$ are connected by tunneling that does not involve net charge transfer to or from the superconductor, yielding zero quantum capacitance. This behavior appears in Fig.~\ref{fig:2}b, where only the even state has a finite signal near $\delta=0$. In addition, the contrast between even and odd parity states reaches a maximum at this point which scales as $\delta C_Q \propto \Delta^{-1}$~\cite{liu2023fusion}. The resulting parity-dependent frequency shift of the resonator is detected in $\Delta \SM$, providing a superconducting analog to reflectometry-based sensing in conventional double dots~\cite{vigneau2023probing}. In our case, the signal arises from Cooper-pair splitting rather than interdot tunneling.

Figure~\ref{fig:2}c shows a measurement of $\Delta \SM$ as the chemical potentials of the two QDs are varied, with weak interdot coupling. The most prominent feature is a strong anti-diagonal line through the joint charge degeneracies, where $|00\rangle$ and $|11\rangle$ form an equal superposition and the total charge is most susceptible to variations of $\delta$. This results in a pronounced signal that we attribute to quantum capacitance, in agreement with Fig.~\ref{fig:2}b. Additional horizontal and vertical bands appear when only one QD is near resonance. These arise from hybridization between a single dot level and the lowest-energy state in the hybrid segment, generating finite quantum capacitance without interdot coupling. Their widths reflect the energy of this state $E_\mathrm{ABS}$, with maximum signal at resonance. Our theoretical model (Fig.~\ref{fig:2}d; see Methods) qualitatively reproduces these features, treating the hybrid segment as a spinful resonant level coupled to a superconductor in the presence of spin-orbit coupling. The energy diagrams in Fig.~\ref{fig:2}e illustrate the processes behind the observed signals. The band visibility depends on QD-hybrid coupling strength, while the joint resonance signal reflects interdot coupling.  Figure~\ref{fig:ed_hybrid_csds} shows how tuning the hybrid electrochemical potential via $\VH$ modifies this behavior. These features help identify suitable joint degeneracies for time-domain parity readout, as discussed next.

\section*{Time-resolved parity measurements}\label{ParityPMM}
\begin{figure*}[b!]
    \centering
    \includegraphics[width=1\textwidth]{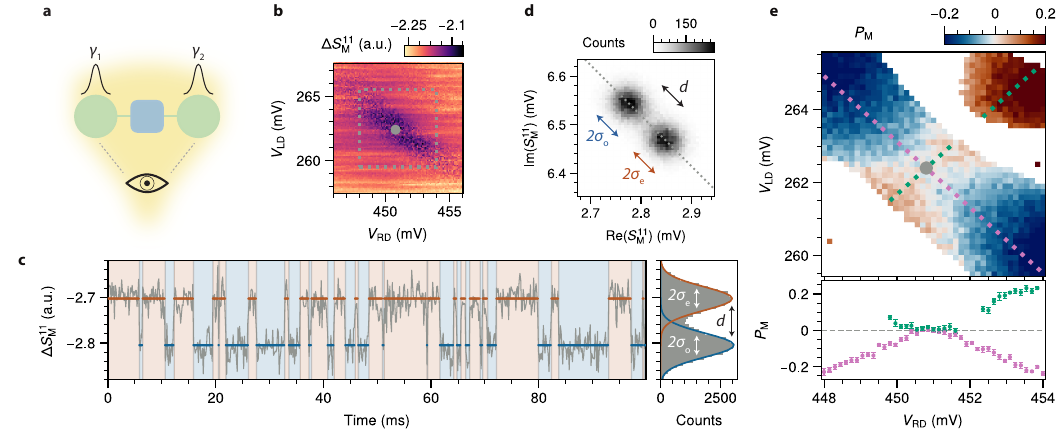}
    \caption{\textbf{Time-resolved parity switching via quantum capacitance}
    \textbf{a,} Illustration of quantum-capacitance readout, which couples to both Majoranas and is sensitive to their parity. 
    \textbf{b,} Charge-stability diagram of QDL and QDR measured through $\Delta \SM$. Random telegraph noise near the anti-diagonal resonance reflects real-time parity switching.
    \textbf{c,} Time trace of $\Delta \SM$ recorded near the center of the charge-stability diagram ($\VQDL = 262.4\,\mathrm{mV}$ and $\VQDR = 450.8\,\mathrm{mV}$, gray dot in panels \textbf{b} and \textbf{e}), showing discrete switching events between parity states. For this trace, an integration time of $\SI{150}{\micro s}$ is used which yields $\mathrm{SNR_M}=\SI{1.958(0.006)}{}$ and readout error of $\SI{6}{\%}$. States are assigned and highlighted in color using a hidden Markov model, which yields $\tau_\mathrm{e} = \SI{1.82(0.05)}{\milli s}$, $\tau_\mathrm{o} = \SI{1.88(0.04)}{\milli s}$. The side panel displays count histograms along the principal component axis. Methods for data acquisition and analysis are described in section \ref{Methods_timetrace}.
    \textbf{d,} Histogram of complex $\SM$ from the time trace in panel \textbf{c,}. The separation $d$ and standard deviation $\sigma_\mathrm{e}$ and $\sigma_\mathrm{o}$ of the distributions are used to calculate the signal-to-noise ratio $\mathrm{SNR} = d/(\sigma_\mathrm{e}+\sigma_\mathrm{o})$. The dotted line indicates the principal component axis along which the data is projected. 
    \textbf{e,} Top: Map of parity polarization $P\mathrm{_M}$ around the joint charge degeneracy inside the dotted gray box of panel \textbf{b}, derived from $\SI{10}{s}$ time traces. Only pixels for which $\mathrm{SNR_M} \geq 0.5$ are included (see also Methods and Fig.~\ref{fig:ed_ectcar}). Bottom: Line cuts of $P\mathrm{_M}$ along the diagonal (green) and anti-diagonal (pink) directions in the top panel, showing approximately quadratic dependence consistent with energy detuning of PMMs from the sweet spot. 
}
    \label{fig:3}
\end{figure*}

We tune the device to a configuration with strong interdot coupling via transport measurements to locate regions where PMM sweet spots are expected (see Methods). We then isolate the device from the left lead and track the relevant quantum dot resonances while compensating for cross-capacitance. In Fig.~\ref{fig:3}, we focus on quantum capacitance measurements as illustrated in Fig.~\ref{fig:3}a. The corresponding charge-stability diagram is shown in Fig.~\ref{fig:3}b. At the center where both dots are on resonance ($\muQDL = \muQDR = 0$), we record a time trace of $\Delta \SM$, which shows discrete switching between two values (Fig.~\ref{fig:3}c). These switching events reflect transitions between even and odd parity ground states. A two-state hidden Markov model (see Methods) identifies these transitions and yields a characteristic switching time of $\tau_\mathrm{avg} = \SI{1.85(0.03)}{\milli s}$, consistent with fits of the power spectral density and the dwell times statistics (Fig.~\ref{fig:ed_power_extra_analysis}). The bimodal histogram (Fig.~\ref{fig:3}d) confirms single-shot parity readout, with optimized integration time of $\SI{150}{\micro s}$ giving a $\SI{6}{\%}$ minimal readout error and achieving a signal-to-noise ratio $\mathrm{SNR_M} = 1$ in $\SI{40}{\micro s}$ (see Fig.~\ref{fig:ed_power_extra_analysis}). The absence of the excited states suggests fast relaxation within each parity sector, consistent with nanosecond timescales for double quantum dots~\cite{petta2004manipulation}.

Next, we extract the lifetimes $\tau_\mathrm{e}$ and $\tau_\mathrm{o}$ of the even and odd parity states. From these, we compute the parity polarization $P_\mathrm{M} = (\tau_\mathrm{e} - \tau_\mathrm{o})/(\tau_\mathrm{e} + \tau_\mathrm{o})$. This quantity indicates the probability of finding the system in either parity state: near zero implies both are equally likely, while positive or negative values indicate an imbalance. We measure $P_\mathrm{M}$ across a grid of chemical potentials around the joint charge degeneracy (dotted gray box in Fig.~\ref{fig:3}b), producing the map shown in Fig.~\ref{fig:3}e. The polarization varies with dot energies, reaching near-zero at the center where dot levels align. Line cuts along the diagonal and anti-diagonal directions show a quadratic variation near the center, consistent with previous observations in minimal Kitaev chains where the energy splitting grows quadratically upon detuning. This behavior can be understood by modeling the system as coupled to a hot reservoir of fermions~\cite{Nguyen2023poisoning,hinderling2024parity} (see Methods). In this model, the system switches between parity states with rates that depend on their energy. The state with lower energy lives longer, so the observed polarization reflects the energy difference between the two states. Apart from the switching between QD parity states detected along the anti-diagonal, hybridization of the dot levels with an Andreev bound state in the hybrid segment produces detectable switching in the top-right corner of the map (see Methods).

We test the robustness of this approach by varying $\VH$, which tunes the CAR and ECT strengths between the dots. As shown in Fig.~\ref{fig:ed_ectcar}, the shape of the polarization region evolves with $\VH$, resembling the transition between CAR and ECT  observed in transport measurements. This suggests that polarization mapping can indicate sweet spot conditions even when the device is isolated. For Fig.~\ref{fig:3}, we chose $\VH = 1.665\,\mathrm{V}$, where the polarization approaches zero near the center of the charge-stability diagram, consistent with a PMM sweet spot. Transport measurements after reconnecting the left lead confirm a sweet spot nearby, at $\VH = 1.675\,\mathrm{V}$ (Fig.~\ref{fig:ed_transport}).

\section*{Poor man's Majorana charge neutrality}\label{QCCS}
\begin{figure*}[t!]
    \centering
    \includegraphics[width=1\textwidth]{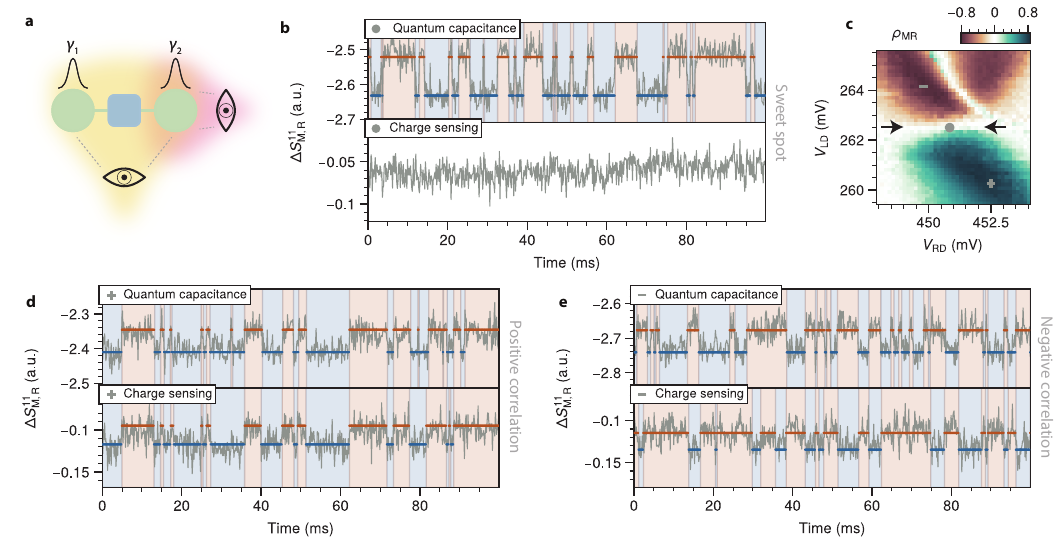}
    \caption{\textbf{Parity readout via quantum capacitance and charge sensing.}
    \textbf{a,} Schematic of the dual readout setup: parity is detected via quantum capacitance  $\Delta \SM$ and charge via a nearby charge sensor $\Delta \SR$.
    \textbf{b,} Time traces recorded near the sweet spot (gray dot in panel c, $\VQDL = \SI{262.5}{\milli V}$ and $\VQDR = \SI{450.8}{\milli V}$). Parity switching is observed in the quantum capacitance signal, whereas the charge sensor does not distinguish parity ($\rho_{\mathrm{MR}} \approx \SI{0.03}{})$. 
    \textbf{c,} Map of the Pearson correlation coefficient $\rho_{\mathrm{MR}}$ across the charge-stability diagram of QDL and QDR, calculated between the state assignment of the two signals. The correlation vanishes near a horizontal line highlighted by black arrows ($\VQDL = \SI{262.5}{\milli V}$), consistent with $t\approx\Delta$.  
    \textbf{d,} Time traces recorded away from degeneracy (gray plus in panel c), where both signals resolve switching events with strong correlation ($\rho_{\mathrm{MR}} \approx \SI{0.85}{}$).
    \textbf{e,} Time traces in the anti-correlated regime (gray minus in panel c). Switching appears in both signals but with opposite sign, reflecting parity inversion without a change in local parity ($\rho_{\mathrm{MR}} \approx \SI{-0.81}{}$). 
}
    \label{fig:4}
\end{figure*}

We next compare two complementary measurements of the Kitaev chain: quantum-capacitance readout via the superconducting lead and charge sensing via a nearby quantum dot (characterized in Fig.~\ref{fig:ed_chargesensor}). Time traces of both signals were recorded simultaneously (Fig.~\ref{fig:4}a). Near the center of the charge-stability diagram, parity switching appears only in the quantum-capacitance signal (Fig.~\ref{fig:4}b). The superconducting probe couples the two PMMs through charge fluctuations on both dots and therefore resolves their parity, whereas the charge sensor remains insensitive because both parity states produce the same average charge on QDR~\cite{leijnse2012parity}.

Detuning the left dot shifts its PMM toward the right, so that both PMMs overlap on QDR~\cite{dvir2023pmm,bordin2025probing} and the charge sensor becomes sensitive to their parity. As such, both the quantum capacitance and charge sensor can detect switching events (Fig.~\ref{fig:4}d). Their signals are strongly correlated, meaning both sensors detect the same transitions. We quantify this using the Pearson correlation coefficient $\rho_\mathrm{MR} \approx \SI{0.85}{}$, where $\rho_\mathrm{MR} = +1$ indicates perfect correlation and $\rho_\mathrm{MR} = -1$ perfect anti-correlation. We attribute imperfect correlations to limited SNR outside optimal operating regimes (see Methods).

Figure~\ref{fig:4}c shows a full map of the correlation coefficient, which is strongest where both sensors clearly detect parity switching and vanishes where either sensor is insensitive. Interestingly, the sign of the correlation flips between the top and bottom halves of the map. This occurs because the global parity of the ground state changes as QDL crosses its charge transition, while the local parity of QDR remains fixed. The quantum capacitance reflects this change, since it senses the combined state of both dots, while the charge sensor responds only to the charge on QDR. As a result, the two measurements respond in opposite ways to the same parity-switching events, leading to anti-correlation (Fig.~\ref{fig:4}e).

At the center of the charge-stability diagram, the correlation coefficient crosses through zero. Here, the charge sensor becomes insensitive to the parity state because both parity branches produce the same average charge on QDR. In this situation, switching between even and odd states no longer causes a detectable change in the charge sensing signal. We predict that this condition occurs along the line $\muQDL = \frac{\Delta - t}{\Delta + t}\,\muQDR$, which becomes horizontal ($\mu_\mathrm{LD} = 0$) at the PMM sweet spot ($t = \Delta$), as explained in detail in Methods. In our data, the zero-correlation line (black arrows in Fig.~\ref{fig:4}c) aligns with this prediction. At other $\VH$ values, the line tilts, consistent with detuning from the sweet spot. In addition, the width of this line reflects the strength of the couplings and, by extension, the gap to the excited states of the chain (see Fig.~\ref{fig:ed_ectcar}, Fig.~\ref{fig:ed_weakcoupling} and Fig.~\ref{fig:ed_sweetspot2} for experimental examples and Methods and Fig.~\ref{fig:ed_theory} for theory).

\section*{Discussion and conclusion}\label{Disc}
To implement parity readout in a minimal Kitaev chain, we combined complementary measurement strategies of charge sensing, parity polarization, and transport to guide device tuning. These tools consistently indicate operation near the PMM sweet spot, though readout remains sensitive to parity even away from degeneracy. In future qubit implementations, any residual energy splitting $E_\mathrm{M} = |t - \Delta|$ would manifest as coherent parity oscillations, allowing optimal fine-tuning via Ramsey-type spectroscopy~\cite{tsintzis2024majorana,luethi2024perfect,pan2025rabi}.

The readout technique introduced here is designed for integration into qubit architectures. In the envisioned geometry of two Kitaev chains connected via a shared superconductor, each chain contributes a parity-dependent quantum capacitance, as demonstrated in Fig.~\ref{fig:ed_pmmqubit}. This enables the single-shot discrimination of the qubit basis states and detects leakage from the computational subspace~\cite{pan2025rabi}. The method also applies to longer chains, where detuning all but two dots effectively reduces the system to a minimal chain during readout. Additionally, we observe that the resonator drive can bias the parity distribution (Fig.~\ref{fig:ed_power_extra_analysis}a,b), suggesting a path toward parity initialization via dynamical polarization~\cite{wesdorp2023dynamical}.

This work overcomes a key challenge in realizing Majorana-based qubits by enabling fast, high-fidelity, single-shot parity readout. While similar functionality has recently been demonstrated using interferometric techniques in extended hybrid devices~\cite{aghaee2025interferometric,aghaee2025distinct}, those approaches require significant alterations to the device layout and introduce considerable control overhead. In contrast, our method integrates seamlessly with existing device elements, enabling parity readout without added operational complexity. Simultaneous charge sensing verifies charge-neutrality, a Majorana signature inaccessible to interferometry. With this capability in place, gate-based control methods become directly applicable. This positions the minimal Kitaev chain as a fully operational platform for time-domain experiments on Majorana fusion, braiding, and computation.

%TC:ignore
\section*{Data Availability and Code Availability}
The raw data measured on the device presented in this work, the data processing and plotting code, and the code used for the theory calculations are available at \url{https://doi.org/10.4121/227fd419-fded-4a96-ab62-421a0cd57fa5}.

\section*{Acknowledgements}
This work has been supported by the Dutch Organization for Scientific Research (NWO), Microsoft Corporation Station Q, the Spanish Ministry of Science (grant No. PID2021-125343NB-I00) and the Horizon Europe Framework Program of the European Commission through the European Innovation Council Pathfinder Grant No. 101115315 (QuKiT). We thank O.W.B. Benningshof and J.D. Mensingh for technical assistance with the cryogenic electronics. We thank F.K. Malinowski for contributing to design of the experiment. We thank C.X. Liu and S.L.D. ten Haaf for their input on the manuscript. We thank T. Dvir, A. Lombardi, V.P.M. Sietses, F.J. Bennebroek Evertsz', M. Wimmer, J.D. Torres Luna, S. Miles, S. Goswami, S. Roelofs, D. Joshi, and A.R. Akhmerov for fruitful discussions. We thank J.M. Hornibrook and D.J. Reilly for providing the frequency multiplexing chips. We thank S. Gazibegovic for contribution to nanowire growth.

\section*{Author contributions}
The sample was fabricated by N.v.L. and B.R. Measurements were performed by N.v.L., F.Z. and T.V.C. The experiment was designed by N.v.L., F.Z. and G.W. The data was analyzed by N.v.L. and F.Z. The experimental setup was designed and implemented by N.v.L., F.Z., B.R., G.W., and D.v.D. The manuscript was prepared by N.v.L. and F.Z. with input from all authors. G.W., A.B., D.v.D., Y.Z., W.D.H. and G.P.M. contributed to understanding and interpretation of the data through regular discussions. The project was supervised by L.P.K. Modeling and simulations of the system were done by G.O.S. and R.A. The InSb nanowire growth was performed by G.B. and E.P.A.M.B.

\section*{Competing interests}
The authors declare no competing interests.
\clearpage

\section*{Extended data}

\setcounter{figure}{0}
\renewcommand{\thefigure}{ED\arabic{figure}}
\renewcommand{\theHfigure}{ED\arabic{figure}}

\begin{figure*}[ht!]
    \centering
    \includegraphics[width=1\textwidth]{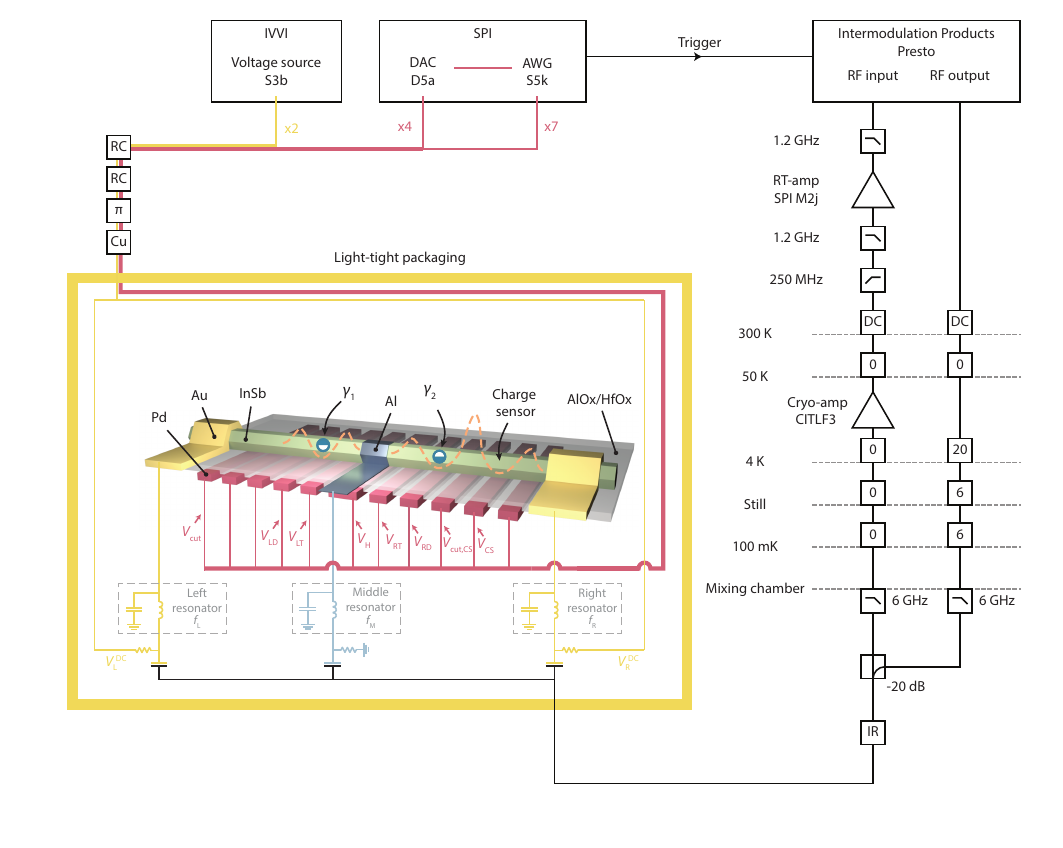}
    \caption{\textbf{Measurement setup.}
    Detailed wiring diagram for the DC and RF lines used to measure the device. To reduce the generation of non-equilibrium quasiparticles in the superconducting film we use IR filters for the RF lines. Additionally, light-tight packaging encloses the device. Finally, we note that an additional resonator was connected to the left normal lead, but it was not used.
}
    \label{fig:ed_measurementsetup}
\end{figure*}

\begin{figure*}[ht!]
    \centering
    \includegraphics[width=1\textwidth]{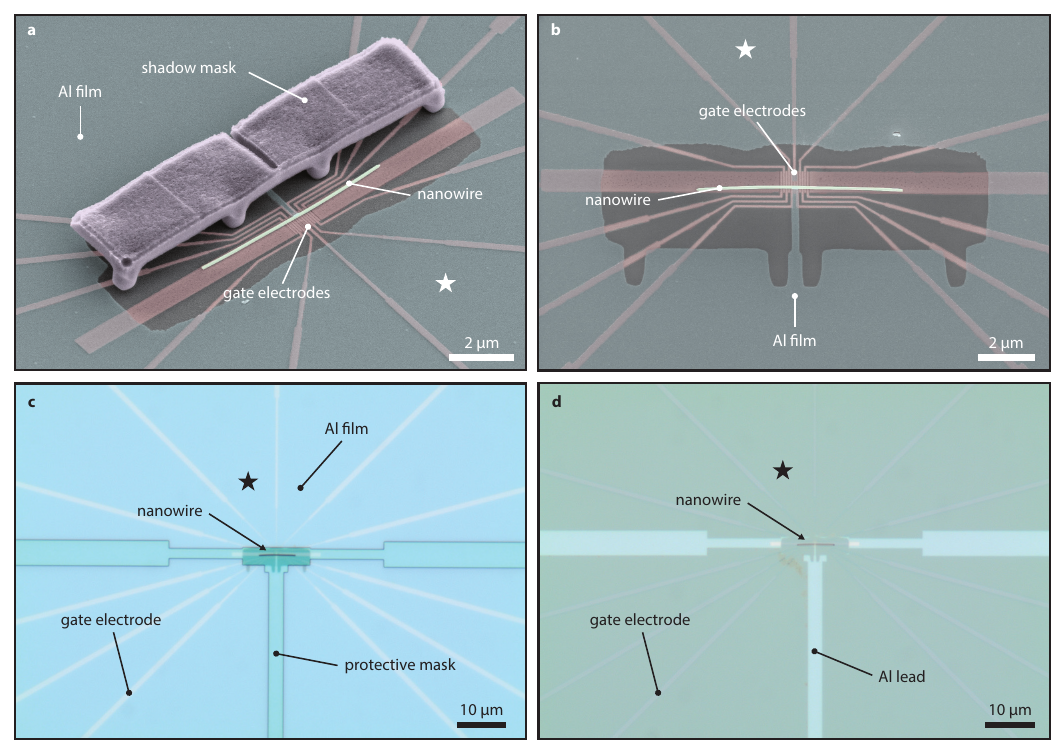}
    \caption{\textbf{Shadow lithography with masks grown on-chip.}
    \textbf{a,} False-colored tilted scanning electron micrograph of a nanowire (InSb, shown in green) on top of the gate electrode array (Pd, shown in red) after deposition of the superconductor (Al, shown in blue). The shadow mask (HSQ, shown in purple) has a narrow slit in the center through which the Al strip that contacts the nanowire is evaporated, at an angle of $30^\circ$ with respect to the substrate. A star is placed at approximately the same location in this and the next panels to help identify the orientation of the substrate. 
    \textbf{b,} False-colored top-view scanning electron micrograph of the same nanowire as in panel \textbf{a}, after the shadow mask has been mechanically removed using a nanomanipulator.
    \textbf{c,} Optical image of the PMMA mask which covers the nanowire and three leads towards the device. Image taken after patterning and development of the mask, but before etching of the Al film using Transene type D.
    \textbf{d,} Optical image of the device after etching of the Al film and stripping of the protective mask. The central Al lead connects the nanowire to a W pad which facilitates wire bonding (not shown here). 
}
    \label{fig:ed_fab}
\end{figure*}

\begin{figure*}[ht!]
    \centering
    \includegraphics[width=1\textwidth]{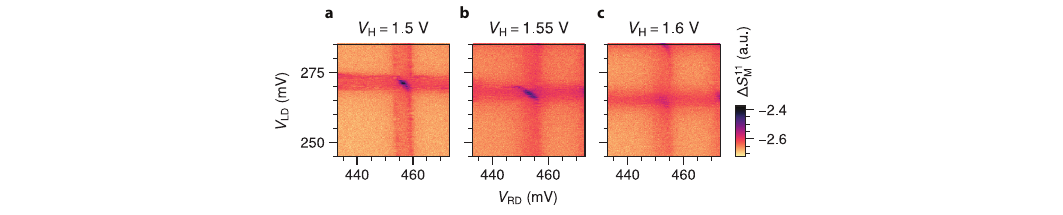}
    \caption{\textbf{Evolution of quantum capacitance with hybrid gate voltage $\VH$}
    \textbf{a--c} Quantum capacitance charge-stability diagrams measured using the middle resonator while the device is isolated from the normal leads. Each panel corresponds to a different setting of the hybrid gate voltage $\VH$, which tunes the energy of the Andreev bound state in the superconducting segment. As $\VH$ increases, we observe a progressive weakening of the signal along the interdot detuning axis. This behavior is attributed to an increase in the strength of CAR between the quantum dots, mediated by the lowering of the Andreev bound state energy. Since quantum capacitance is inversely proportional to CAR strength, a stronger CAR leads to reduced signal contrast. This trend provides a qualitative probe of the CAR amplitude and helps identify regions where $t \approx \Delta$, as required for the PMM sweet spot.
}
    \label{fig:ed_hybrid_csds}
\end{figure*}

\begin{figure*}[ht!]
    \centering
    \includegraphics[width=1\textwidth]{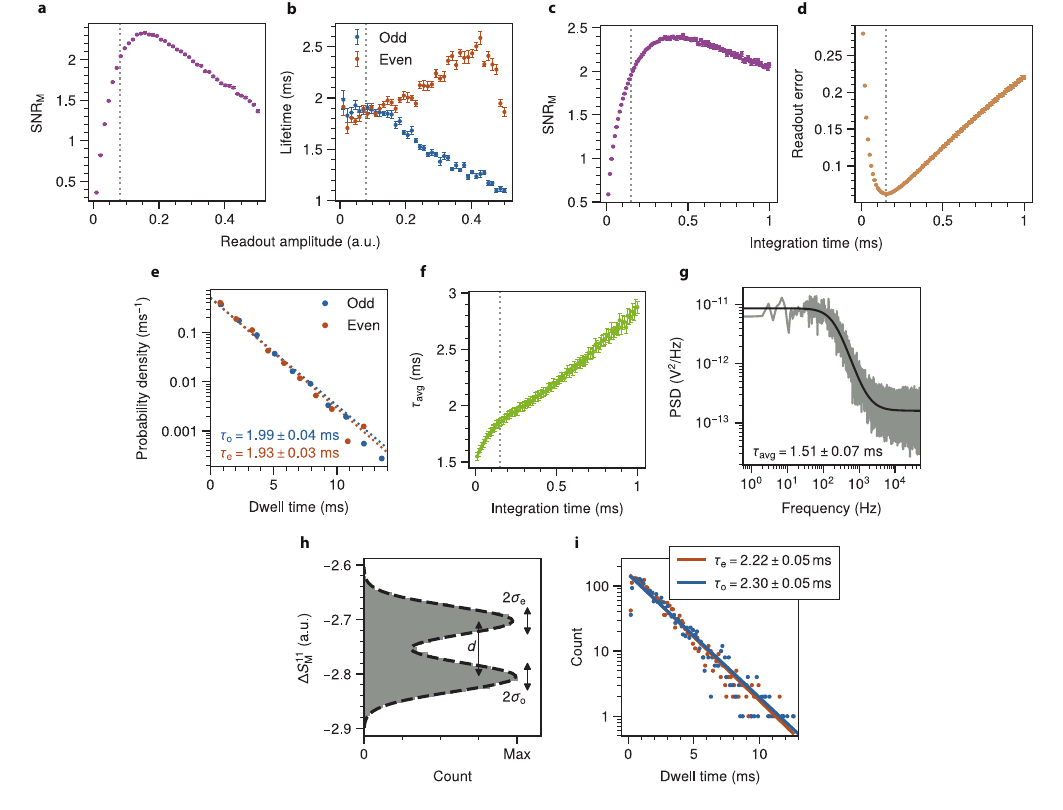}
    \caption{\textbf{Detailed analysis of time-trace performance at the PMM sweet spot.}
    \textbf{a,b,} Signal-to-noise ratio (SNR$_\mathrm{M}$, panel \textbf{a}) and parity lifetimes (panel \textbf{b}) as a function of readout amplitude. All data were taken with the same gate settings and $\SI{150}{\micro s}$ integration time as in Fig.~\ref{fig:3}d. The gray dotted line indicates the amplitude used for Fig.~\ref{fig:3} and Fig.~\ref{fig:4}. Beyond a certain readout amplitude, the SNR decreases and the lifetimes become parity-dependent, consistent with drive-induced modulation along the common-mode axis $\delta$ that preferentially stabilizes the even ground state. This observation suggests that the system can be initialized in the even state by driving the middle resonator with a high amplitude. The time traces in panels \textbf{a},\textbf{b} were averaged in time bins of $\SI{160}{\micro s}$ (instead of $\SI{150}{\micro s}$) to ensure that the averaging time is an exact multiple of the sampling rate ($\SI{50}{\kilo Hz}$).
    \textbf{c,d,} SNR$_\mathrm{M}$ (\textbf{c}) and readout error (\textbf{d}) of the time trace presented in Fig.~\ref{fig:3} versus integration time. The readout error was estimated as $\left[1-e^{-\tau_\mathrm{bin}/\tau_\mathrm{avg}}\mathrm{erf}\left(\mathrm{SNR_{M}(\tau_\mathrm{bin})}/\sqrt{2}\right)\right]/2$ \cite{aghaee2025interferometric}, where $\tau_\mathrm{bin}$ is the integration time and $\tau_\mathrm{avg} = \SI{1.85(0.03)}{ms}$. The gray dotted line indicates the integration time $\tau_\mathrm{bin} = \SI{150}{us}$ used for all the reported measurements unless otherwise specified.
    \textbf{e,} Histograms of dwell times in the even and odd parity states, normalized by total counts. The dotted lines show fits with the corresponding Poissonian probability densities $e^{-t/\tau_\mathrm{o,e}}/\tau_\mathrm{o,e}$, confirming that the dwell times follow an exponential distribution. $\tau_\mathrm{o,e}$ have been estimated as the mean of the dwell times in the odd and even parity state. Their uncertainty has been estimated as the standard deviation of the mean.
    \textbf{f,} Average switching time, $\tau_\mathrm{avg}$, estimated for varying integration time. As the integration time increases, the estimation of $\tau_\mathrm{avg}$ is systematically biased towards higher values. See Methods section~\ref{Methods_timetrace} for a more detailed discussion.
    \textbf{g,} Power spectral density (PSD) of the time trace using an integration time of $\SI{10}{\micro s}$. By fitting it with a Lorentzian model, we extract an average switching time of $\SI{1.51(0.07)}{\milli s}$. The discrepancy between this estimate and those obtained with a hidden Markov model or exponential fitting further suggests that the long integration time needed to obtain high SNR could systematically bias the estimation of $\tau_\mathrm{avg}$. See Methods section~\ref{Methods_timetrace} for a more detailed discussion.
    \textbf{h,i,} Analysis of the time trace of Fig.~\ref{fig:3}c using the analysis methods and code adapted from Ref.~\cite{aghaee2025interferometric}. In panel \textbf{h}, the histogram is fit to a weighted sum of Gaussian distributions. The extracted $\mathrm{SNR} = d/(\sigma_\mathrm{e}+\sigma_\mathrm{o})=\SI{1.93}{}$ is comparable with the hidden Markov model analysis. In panel \textbf{i}, the dwell time distributions are extracted using a Gaussian mixture model and a thresholding algorithm. The parity lifetimes are then estimated by fitting the distributions with an exponential model. 
    The estimates are comparable to those obtained with the hidden Markov model analysis.
}
    \label{fig:ed_power_extra_analysis}
\end{figure*}

\begin{figure*}[ht!]
    \centering
    \includegraphics[width=1\textwidth]{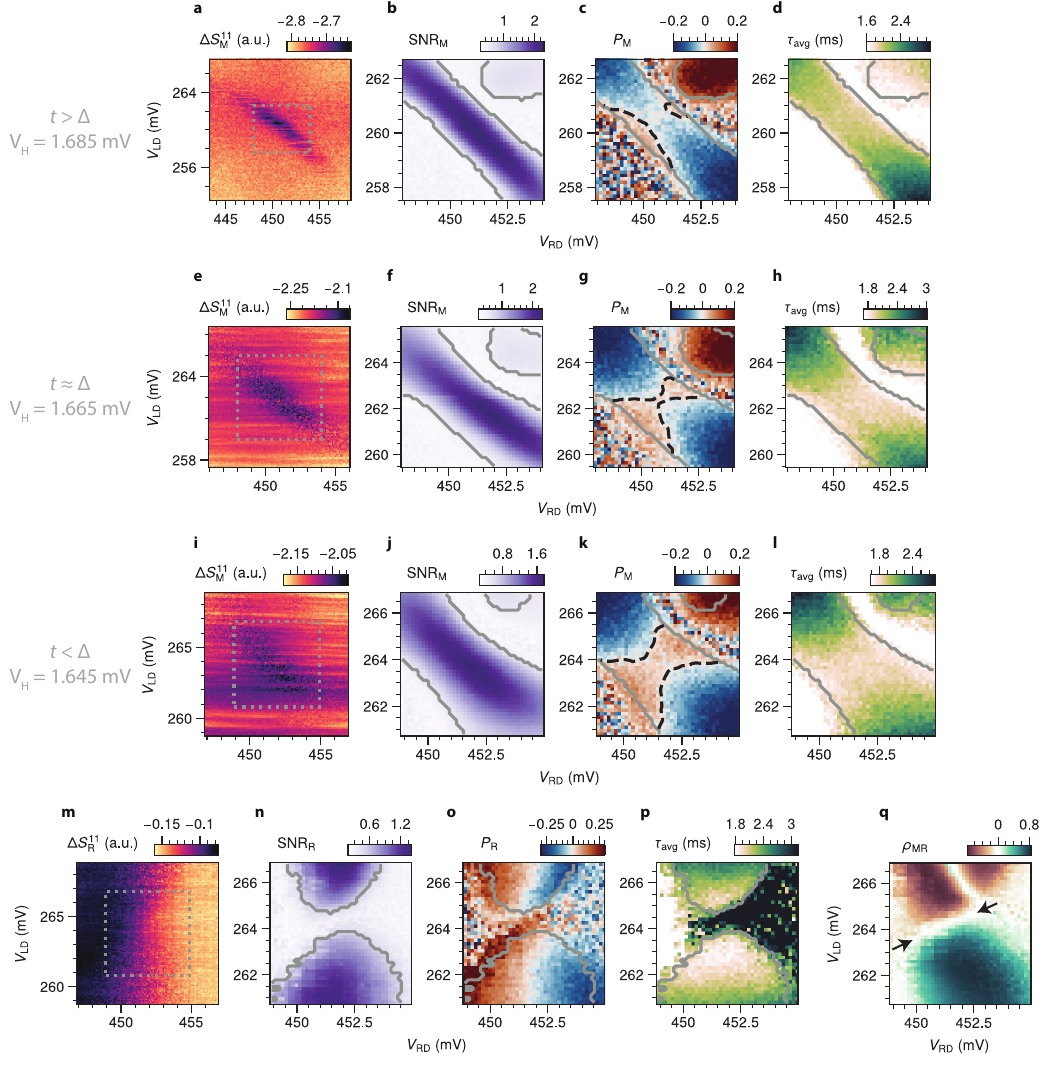}
    \caption{\textbf{Evolution of charge-stability diagrams and switching behavior across the ECT-to-CAR transition.}
    \textbf{a--d,} Measurements at $\VH = \SI{1.685}{V}$, where $t>\Delta$. 
    The charge-stability diagram measured via quantum capacitance (\textbf{a}) shows strong signal along the detuning axis. The gray dotted box marks the region where 10\,s time traces were measured to estimate (Methods section~\ref{Methods_timetrace}) the signal-to-noise ratio (SNR$_\mathrm{M}$, panel \textbf{b}), parity polarization ($P_\mathrm{M}$, panel \textbf{c}), and average switching time ($\tau_\mathrm{avg}$, panel \textbf{d}). In these panels, the gray contours indicate where $\mathrm{SNR_M}=0.5$. In panel \textbf{c}, a black dashed line highlights the contour where $P_\mathrm{M} = 0$, calculated after applying a Gaussian filter with standard deviation of 1 pixel to reduce noise. This contour resembles an avoided crossing along the detuning axis, characteristic of the condition $t>\Delta$. 
    \textbf{e--h,} Same measurements at $\VH = \SI{1.665}{V}$, where $t \approx \Delta$ near the PMM sweet spot. In this case, the parity polarization map (panel \textbf{g}) shows no avoided crossing. Instead, regions of positive and negative parity polarization alternate around the center of the charge stability diagram, indicating that CAR and ECT are balanced. This is the dataset used in the main text.
    \textbf{i--l,} Same at $\VH = \SI{1.645}{V}$, where CAR dominates ($t < \Delta$). Here, the polarization map (panel \textbf{k}) shows that the avoided crossing reappears but now along the common-mode axis, reversing the pattern seen in \textbf{c}. This indicates that the system transitioned to a regime where $t<\Delta$.
    \textbf{m--p,} Corresponding data from panels \textbf{i--l} using the charge sensor, recorded via the right resonator. The SNR$_\mathrm{R}$ vanishes at the center of the charge-stability diagram, when the QDL is on resonance ($\VQDL \approx \SI{264}{\milli V}$). This indicates that a charge measurement of QDR is not sufficient to determine the parity of the system.
    \textbf{q,} Map of the Pearson correlation coefficient $\rho_\mathrm{MR}$ between quantum capacitance and charge sensor signals in the $t<\Delta$ regime. The sign change of $\rho_\mathrm{MR}$ (black arrows) no longer follows a horizontal line. This is consistent with CAR dominating over ECT, as explained in more detail in Methods section~\ref{Methods:Theory}.
}
    \label{fig:ed_ectcar}
\end{figure*}

\begin{figure*}[ht!]
    \centering
    \includegraphics[width=1\textwidth]{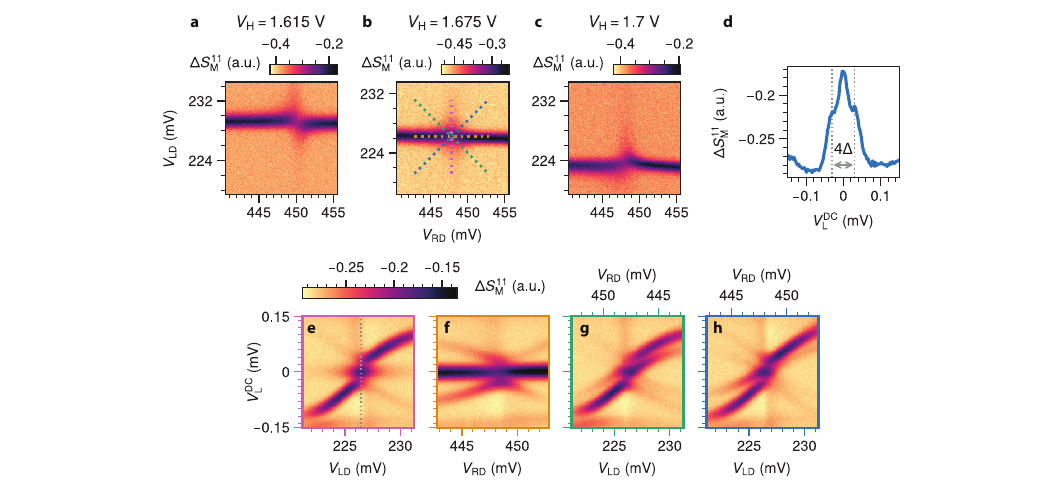}
    \caption{\textbf{Transport measurement of the PMM sweet spot using RF reflectometry.}
    \textbf{a--c,} Zero-bias charge-stability diagrams for different values of $\VH$, used to tune the relative strength of elastic co-tunneling (ECT) and crossed Andreev reflection (CAR). In all panels, the left lead is operated as a tunnel probe, while the right side is blocked by the charge sensor. As a result, transport occurs predominantly through QDL, and visibility of the QDR resonance is suppressed. The orientation of the avoided crossing in panel \textbf{a} indicates that $t<\Delta$. As $\VH$ increases, the avoided crossing first disappears when the sweet spot $t\approx \Delta$ (\textbf{b}) is approached. Further increasing $\VH$ opens the avoided crossing in the opposite direction, indicating that $t>\Delta$ (\textbf{c}). The dotted lines in panel \textbf{b} indicate the axes along which the spectra of panels \textbf{e--h} were measured.
    \textbf{d,} Tunnel spectroscopy varying $\VLDC$ at the center of the charge-stability diagram of panel \textbf{b}. The spectrum exhibits a zero-bias peak separated from two excited states by a gap of $\approx \SI{30}{\micro eV}$, consistent with PMMs in a minimal Kitaev chain. This linecut was extracted along the gray dotted line of panel \textbf{e}.
    \textbf{e, f,} Spectra measured while varying $\VLDC$ and the electrochemical potential of the left (\textbf{e}) or right (\textbf{f}) QD. The stable zero-bias peak indicates robustness against local perturbations.
    \textbf{g, h,} Spectra measured while varying $\VLDC$ and the electrochemical potential of both QDs along the common-mode (\textbf{g}) and detuning (\textbf{h}) axis. The zero-bias peak splits quadratically, illustrating limited protection against global perturbations.  
}
    \label{fig:ed_transport}
\end{figure*}

\begin{figure*}[ht!]
    \centering
    \includegraphics[width=1\textwidth]{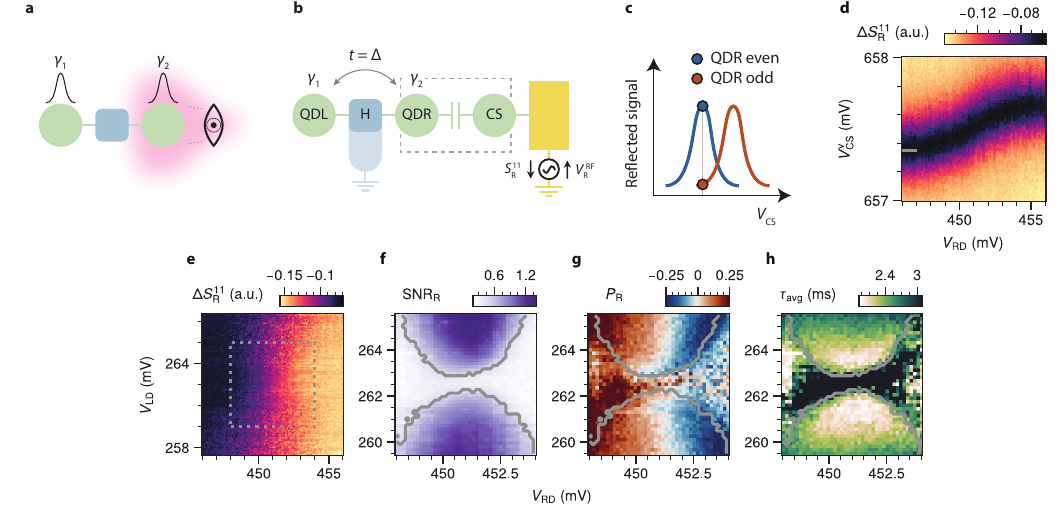}
    \caption{\textbf{Parity readout using a nearby charge sensor.}
    \textbf{a,} Illustration showing that a probe which does not couple the two Majoranas to each other cannot access the parity $i\gamma_1\gamma_2$, and is thus insensitive to the stored quantum information
    \textbf{b,} Schematic of the charge sensing readout. A nearby quantum dot (CS) is capacitively coupled to QDR but isolated by a high tunnel barrier that prevents direct tunneling between them. A lead with a resonator is coupled to the CS for fast single-lead RF reflectometry~\cite{persson2010excess}.
    \textbf{c,} Principle of charge sensing: changes in the charge on QDR alters the local electrostatic environment, shifting the CS resonance and modulating the reflected signal $\Delta \SR$. 
    \textbf{d,} Charge-stability diagram of QDR and CS measured via $\Delta \SR$, with QDL held on resonance ($\VQDL = \SI{263}{\milli V}$). To correct for cross-capacitance, the voltage on $\VCS$ is adjusted during $\VQDR$ sweeps, resulting in an effective compensated axis $\VCS^\mathrm{v}$ that is a linear combination of $\VQDR$ and $\VCS$. 
    \textbf{e,} Charge-stability diagram of QDL and QDR with $\VCS^\mathrm{v}$ fixed (gray mark in panel \textbf{d}). Outside the QDL resonance, QDR shows fixed charge when its energy is above the hybrid gap $E_\mathrm{ABS}$. Below this gap, parity switching is visible as signal noise. Near the QDL resonance ($\VQDL = \SI{263}{\milli V}$), the region of visible switching is reduced when QDL becomes lower in energy than QDR.
    \textbf{f--h,} The signal-to-noise ratio (SNR$_\mathrm{R}$, panel \textbf{f}), parity polarization ($P_\mathrm{R}$, panel \textbf{g}), and average switching time ($\tau_\mathrm{avg}$, panel \textbf{h}), obtained from time traces measured in the dotted gray box of panel \textbf{e}. In these panels, the gray contours indicate where $\mathrm{SNR_R}=0.5$. The SNR$_\mathrm{R}$ vanishes at the center of the diagram (where $\VQDL \approx 262.5,\mathrm{mV}$), indicating that the charge sensor is insensitive to the parity states at this point. This dataset was acquired simultaneously with the quantum capacitance measurements in Fig.~\ref{fig:3}e, and used to compute correlations in Fig.~\ref{fig:4}.
}
    \label{fig:ed_chargesensor}
\end{figure*}

\begin{figure*}[ht!]
    \centering
    \includegraphics[width=1\textwidth]{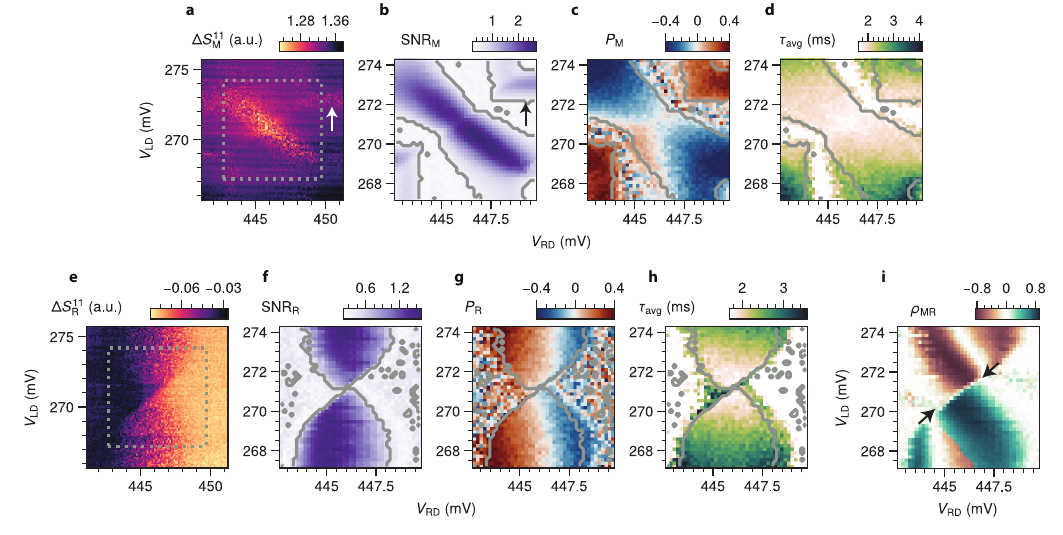}
    \caption{\textbf{Time traces with weakly coupled QDs.}
    \textbf{a--d,} Charge-stability diagram measured via quantum capacitance (panel \textbf{a}) in a weak interdot coupling regime ($\VH = \SI{1.445}{V}$). The gray dotted box marks the region where time traces were measured to estimate the signal-to-noise ratio (SNR$_\mathrm{M}$, panel \textbf{b}), parity polarization ($P_\mathrm{M}$, panel \textbf{c}), and average switching time ($\tau_\mathrm{avg}$, panel \textbf{d}). We infer the weak coupling from the presence of signal when a QD is aligned to the lowest energy state in the hybrid, as highlighted by the arrows in panels \textbf{a,b} and discussed in Fig.~\ref{fig:2}. In these panels, the gray contours indicate where $\mathrm{SNR_M}=0.5$. The probing frequency for the middle resonator was $\SI{341}{\mega Hz}$ instead of $\SI{340.8}{\mega Hz}$ as in the rest of the text. The time traces in this figure were averaged in time bins of $\SI{200}{\micro s}$ (instead of $\SI{150}{\micro s}$) to ensure that the averaging time is an exact multiple of inverse of the sampling rate ($\SI{10}{\kilo Hz}$).
    \textbf{e--h,} Same measurements via charge sensing. Charge-stability diagram of QDL and QDR (panel \textbf{e}). Outside the QDL resonance, QDR shows fixed charge when its energy is above the hybrid gap $E_\mathrm{ABS}$. Below this gap, parity switching is visible as signal noise. Near the QDL resonance, the region of visible switching is reduced when QDL becomes lower in energy than QDR. The region where even and odd states are indistinguishable is narrower than in Fig.~\ref{fig:ed_chargesensor} and Fig.~\ref{fig:ed_sweetspot2}, consistent with small coupling between QDs (see Methods section~\ref{Methods:Theory}). 
    \textbf{i,} Pearson correlation coefficient between time traces measured with quantum capacitance and charge sensing. Similarly to Fig.~\ref{fig:ed_ectcar}q, the transition region between positive and negative correlation (black arrows) is tilted. Additionally, this transition is more sharply defined, in line with the low coupling (see Methods section~\ref{Methods:Theory}).
}
    \label{fig:ed_weakcoupling}
\end{figure*}

\begin{figure*}[ht!]
    \centering
    \includegraphics[width=1\textwidth]{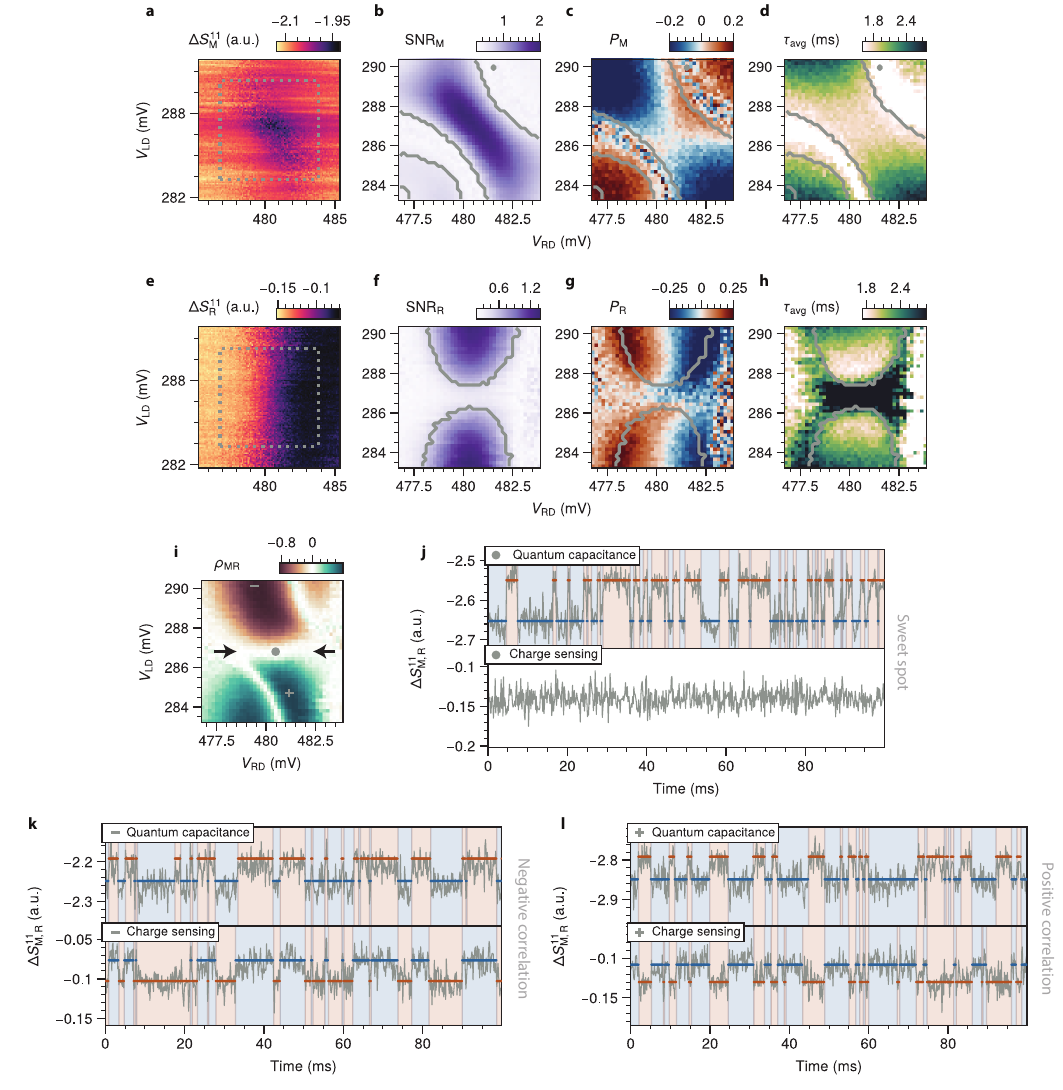}
    \caption{\textbf{ PMM sweet spot reproduction with different QD resonances.}
    \textbf{a--d,} Charge-stability diagram at $\VH=\SI{1.625}{V}$ measured via quantum capacitance (panel \textbf{a}), with the corresponding signal-to-noise ratio (SNR$_\mathrm{M}$, panel \textbf{b}), parity polarization ($P_\mathrm{M}$, panel \textbf{c}), and average switching time ($\tau_\mathrm{avg}$, panel \textbf{d}) for different QD resonances where $t\approx\Delta$. Panels \textbf{b--d} were measured inside the gray dotted box of panel \textbf{a}. In these panels, the gray contours indicate where $\mathrm{SNR_M}=0.5$.
    \textbf{e--h,} Same measurements via charge sensing. The charge-stability diagram of QDL and QDR is shown in panel \textbf{e}. Outside the QDL resonance, QDR shows fixed charge when its energy is above the hybrid gap $E_\mathrm{ABS}$. Below this gap, parity switching is visible as signal noise. Near the QDL resonance ($\VQDL = \SI{286.8}{\milli V}$), the region of visible switching is reduced when QDL becomes lower in energy than QDR. The signal-to-noise ratio (SNR$_\mathrm{R}$, panel \textbf{f}), parity polarization ($P_\mathrm{R}$, panel \textbf{g}), and average switching time ($\tau_\mathrm{avg}$, panel \textbf{h}) are obtained from time traces measured in the dotted gray box of panel \textbf{e}. In these panels, the gray contours indicate where $\mathrm{SNR_R}=0.5$. The region where SNR$_\mathrm{R}$ vanishes near the center of the diagram is wider than for Fig.~\ref{fig:ed_chargesensor} and Fig.~\ref{fig:ed_weakcoupling}, indicating a stronger coupling.
    \textbf{i,} Pearson correlation coefficient between time traces measured with quantum capacitance and charge sensing. The black arrows indicate the region where the charge sensor is insensitive to parity switching events, resulting in vanishing correlation between time traces measured with quantum capacitance and charge sensing. Like for Fig.~\ref{fig:4}, this line is horizontal ($\mu_\mathrm{LD} = 0$) which occurs only at the PMM sweet spot ($t = \Delta$).
    \textbf{j--l,} Examples of uncorrelated (\textbf{j}), negatively correlated (\textbf{k}), and positively correlated (\textbf{l}) time traces. The marks in panel \textbf{i} indicate the specific gate values where these time traces were measured. Note that for panel \textbf{k} and \textbf{l} the reflected signal is, respectively, positively and negatively correlated, opposite to the state assignment (see Methods section~\ref{Methods_state_assignment} for more details). 
}
    \label{fig:ed_sweetspot2}
\end{figure*}

\begin{figure*}[ht!]
    \centering
    \includegraphics[width=1\textwidth]{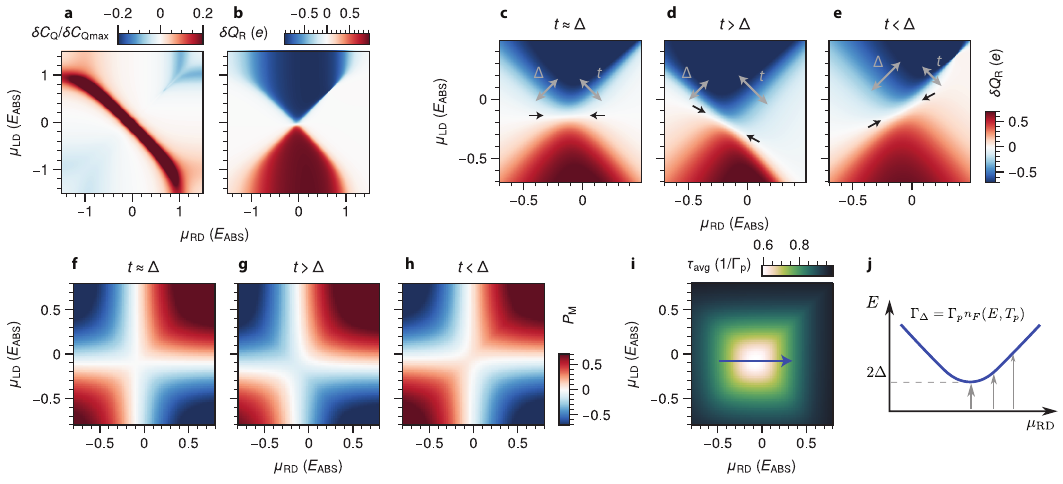}
    \caption{\textbf{Theory model of quantum capacitance, charge sensing, and quasiparticle poisoning dynamics.}
    \textbf{a,b,} Quantum capacitance (\textbf{a}) and right QD charge difference (\textbf{b}) for $t_L=t_R=0.3E_\mathrm{ABS}$, $\theta=\pi/8$, and $\mu_C=0.75\Delta_P$, corresponding to a PMM sweet spot. Red and blue areas indicate if the odd or even state yields the largest signal, and white where the states are indistinguishable.
    \textbf{c--e,} Zooms of QD charge difference with $t_L=t_R=0.5E_\mathrm{ABS}$, $\theta=\pi/8$, and $\mu_C = 0.75\Delta_P$ (\textbf{c}), $\mu_C =2.5\Delta_P$ (\textbf{d}), and $\mu_C = 0.3\Delta_P$ (\textbf{e}). Here, various tilts of the $\mu_\mathrm{LD}=\frac{\Delta-t}{\Delta+t}\mu_\mathrm{RD}$ line are shown as emphasized by the black arrows. Gray arrows indicate the width of the transitions, set by $\Delta$ and $t$.
    \textbf{f--h,} Parity polarization maps for $t_L=t_R=0.4E_\mathrm{ABS}$, $\theta=\pi/8$, $T_p = 0.2\Delta_P$, with $\mu_C = 0.75\Delta_P$ (\textbf{f}), $\mu_C =2.5\Delta_P$ (\textbf{g}), and $\mu_C=0.3\Delta_P$ (\textbf{h}). These maps match the experimental behavior observed in Fig.~\ref{fig:ed_ectcar}.
    \textbf{i,} Average parity lifetime $\tau_{\mathrm{avg}}=1/(\Gamma_\mathrm{e\rightarrow o}+\Gamma_\mathrm{o\rightarrow e})$ for same parameters as panel \textbf{f}.
    \textbf{j,} Schematic of the poisoning process, $\Gamma_\mathrm{\Delta}$, which decreases the lifetime in the center of \textbf{i} via poisoning to the excited states. The x-axis corresponds to the blue arrow in \textbf{i}. Details on the model can be found in Methods section~\ref{Methods:Theory}, with calculations on the quantum capacitance and charge sensing signals in subsection~\ref{Methods:TheoryQCCS} and state occupation in subsection~\ref{Methods:TheoryOccupation}.
}
    \label{fig:ed_theory}
\end{figure*}

\begin{figure*}[ht!]
    \centering
    \includegraphics[width=1\textwidth]{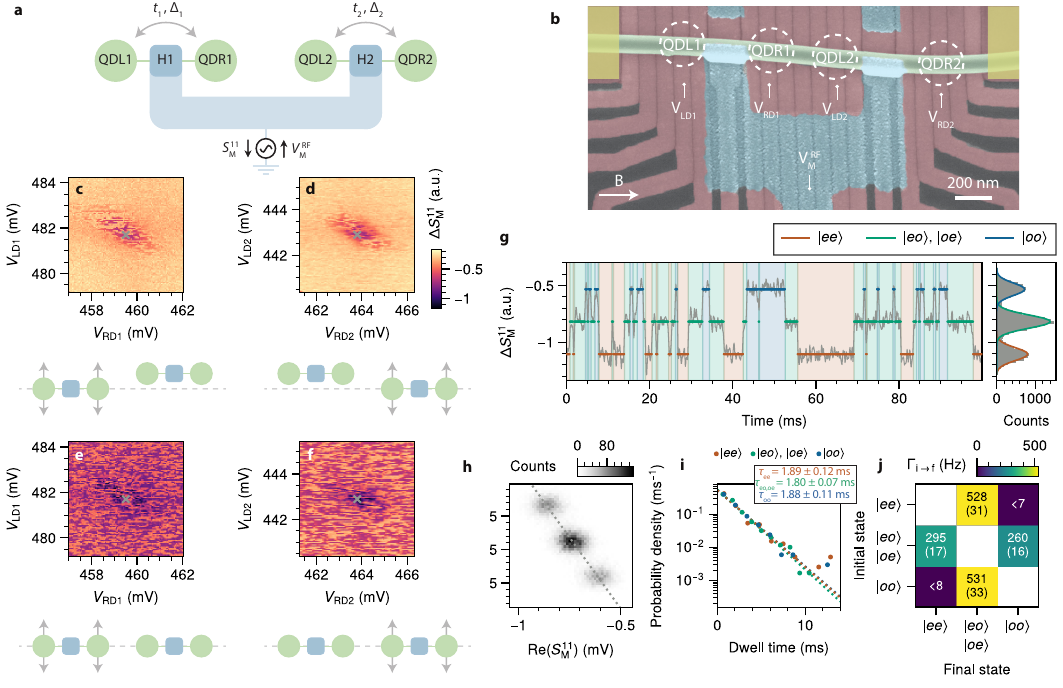}
    \caption{\textbf{Quantum capacitance parity readout of two minimal chains in a qubit geometry.}
    \textbf{a,} Schematic of a device with two minimal Kitaev chains coupled to a common superconductor and an off-chip resonator. Each chain contributes a parity-dependent quantum capacitance: $C_\mathrm{Q}^{1(2)}\sim 1/\Delta_\mathrm{1(2)}$ (if the chain is even), and $C_\mathrm{Q}^{1(2)}=0$ (if the chain is odd), where $\Delta_{1(2)}$ is the CAR coupling of the first (second) chain. The total quantum capacitance for the four possible parity states is $C_\mathrm{Q}^{\mathrm{ee}}\sim1/\Delta_\mathrm{1}+1/\Delta_\mathrm{2}$, $C_\mathrm{Q}^{\mathrm{eo}}\sim1/\Delta_\mathrm{1}$, $C_\mathrm{Q}^{\mathrm{oe}}\sim1/\Delta_\mathrm{2}$, $C_\mathrm{Q}^{\mathrm{oo}}=0$. If the chains are asymmetric ($\Delta_1\neq \Delta_2$), four states can be distinguished. On the other hand, if $\Delta_1 \sim \Delta_2$ the two globally odd states cannot be distinguished. Nevertheless, the two even states can always be distinguished both from each other and from the globally odd state~\cite{pan2025rabi}.
    \textbf{b,} False-colored SEM of a device with two chains. The nanowire is depleted between inner dots QDR1 and QDL2 to prevent inter-chain tunneling. Ohmic contacts (yellow) were added in a second fabrication step. The rightmost quantum dot was shifted by one gate to the right, because the gate closest to the superconductor was too strongly screened to form a suitable tunnel barrier.
    \textbf{c,d,} Charge stability diagram of the left chain, while the QDs of the right chain are off resonance (\textbf{c}) and vice versa (\textbf{d}). In this case, the quantum capacitance readout is sensitive to the parity of only one chain. Parity switching events appear as telegraph noise in the center of the charge stability diagrams, as discussed in Fig.~\ref{fig:2}. For this measurement and the following ones, the two chains were not fine-tuned to the sweet spot $t = \Delta$.
    \textbf{e,f,} Charge stability diagram of the left chain, while the QDs of the right chain are on resonance (\textbf{e}) and vice versa (\textbf{f}). The telegraph noise in the background corresponds to parity switching events of the chain whose QDs are not being detuned. At the center of the charge stability diagram all the QDs are on resonance and more than two-states can be detected. 
    \textbf{g,} 
    Time trace with the QDs of both chains on resonance, taken at the gate settings marked by gray crosses in (c–f). Three levels are resolved, consistent with $\Delta_1 \sim \Delta_2$ where the odd states merge. A Gaussian mixture model (see Methods section~\ref{Methods_timetrace}) identifies the states, with the central state twice as probable. This result demonstrates single-shot readout in the computational basis $\ket{ee}, \ket{oo}$ for a potential PMM qubit, while also enabling detection of leakage into the global odd manifold. A full investigation of parity-qubit operation will be the subject of future work.
    \textbf{h,} Histogram of the complex time trace plotted in \textbf{g}. The one-dimensional time trace was obtained by projecting the complex data onto the principal component axis, drawn here with a dotted line.
    \textbf{i,} Distribution of the dwell times for the three identified states. The good match with an exponential model confirms that the switching processes are Poissonian. The average dwell times are comparable to those measured in the single-chain device.
    \textbf{j,} Transition rates and their standard deviation for the three-state switching process. No direct $\ket{ee} \leftrightarrow \ket{oo}$ were observed. The corresponding matrix entries are the estimated upper bound ($95\%$ CL). This confirms the absence of inter-chain tunneling.}
    \label{fig:ed_pmmqubit}

\end{figure*}

\clearpage
\section*{Methods}\label{Methods}
\renewcommand{\thesection}{}
\renewcommand{\thesubsection}{\Roman{subsection}}
\renewcommand{\thesubsubsection}{\Alph{subsubsection}}

\let \savenumberline \numberline
\def \numberline#1{\savenumberline{#1.}}
\subsection{Device fabrication}\label{Methods_fab}
The device consists of a 95\,nm diameter InSb nanowire~\cite{badawy2019high} contacted by an Al film that induces superconductivity, and is fabricated using a modified shadow-wall lithography technique based on Refs~\cite{heedt2021shadow,mazur2022spin}. 
The base substrate consists of high-resistivity Si with a 100\,nm Si$_3$N$_4$ layer grown via low-pressure chemical vapor deposition. Bottom gates (3\,nm/7\,nm Ti/Pd) were deposited by electron-beam evaporation, with 50\,nm W bond pads added via RF sputtering. A 20\,nm/10\,nm Al$_2$O$_3$/HfO$_2$ gate dielectric was grown using atomic layer deposition.
To pattern the superconductor, a local shadow mask was created near the fine gate array. The substrate was first coated with PMMA (950k A8, $\sim \SI{800}{\nano m}$ thick), exposed using electron-beam lithography (EBL) (500×500\,nm squares), and developed in MIBK:IPA (1:3). A 600\,nm HSQ (FOx-25) layer was then spun on, baked, and patterned via EBL to define the superconductor openings. HSQ was developed in MF-21A at 50°C and PMMA removed in acetone, resulting in suspended masks supported by HSQ pillars.
The nanowire is placed using an optical nanomanipulator, followed by hydrogen-radical cleaning in an electron-beam evaporator. A 17.5\,nm Al layer was deposited at a 30° angle to form the $\sim$180\,nm wide hybrid segment, then oxidized in 200\,mtorr O$_2$ for 5\,minutes. In Fig.~\ref{fig:ed_fab}a, we show an example of a nanowire after deposition of the Al film, together with the shadow mask. 
Following deposition, masks were mechanically removed. The same device after removal of the mask is shown in Fig.~\ref{fig:ed_fab}b. To etch excess Al, PMMA (950k A4) was applied, vacuum-cured, and patterned via EBL to protect the hybrid section (see Fig.~\ref{fig:ed_fab}c). The Al was etched using Transene type D (60\,s at room temperature), rinsed in H$_2$O, and resist stripped in acetone. The result is shown in Fig.~\ref{fig:ed_fab}d. Finally, ohmic contacts (10\,nm/90\,nm Cr/Au) were defined by EBL and deposited via electron-beam evaporation. The device in Fig.~\ref{fig:ed_pmmqubit} was fabricated using the same recipe. Additional fabrication details including substrate preparation, nanowire surface treatment, superconductor growth, and ohmic contact fabrication are described in Ref~\cite{mazur2022spin}.

\subsection{Measurement setup}
\label{Methods_setup}
The sample was measured in an Oxford Triton-400 dilution refrigerator with a base temperature of $\sim20\,\mathrm{mK}$, which is equipped with a 6/1/1\,T vector magnet. The contacts to the nanowire are wire-bonded to LC resonators on a separate frequency-multiplexing chip~\cite{hornibrook2014frequency}. The spiral inductors bonded to the superconductor and right lead have a nominal inductance of $\SI{570}{\nano H}$ and $\SI{150}{\nano H}$, respectively. These resonators enable RF reflectometry via voltages $\VMRF$ and $\VRRF$ at the respective resonance frequencies $f_M \approx \SI{340.8}{\mega Hz}$ and $f_R \approx \SI{704.0}{\mega Hz}$. The superconductor is grounded through a bias tee while DC voltages $\VLDC$ and $\VRDC$ can be applied to the normal leads.
We use a Minicircuits ZUDC20-83-S+ directional coupler to couple the RF input and output lines. The reflected signal is first amplified via a Cosmic Microwave Tech CITLF3 high-electron-mobility transistor (HEMT) at the $\SI{4}{K}$ stage and additionally via a room temperature amplifier built in-house (M2j module for SPI rack). RF signals are generated and demodulated using a Presto from Intermodulation Products. Infrared filters (XMA 2482-5002CL-CRYO), as well as several layers of shielding, are used to reduce noise and protect the sample from radiation that generates non-equilibrium quasiparticles in the superconducting film. 
Voltages applied to gate electrodes are controlled using digital to analog converters built in-house. Additionally, an arbitrary waveform generator built in-house (S5k module for SPI rack) is used to generate sawtooth signals for fast gate sweeps. If their frequency is much smaller than the bandwidth of the DC lines (limited by a second-order RC filter board with $\SI{1}{\kilo Ohm}$ resistors and $\SI{1}{\nano F}$ capacitors), they can be summed on top of the DC voltages for fast RF measurements. Two-dimensional measurement data is then obtained all at once (or line by line) instead of point by point, thus reducing the communication-time overhead. Averaging is performed to increase the quality of the signal. Fig.~\ref{fig:2}c, Fig.~\ref{fig:ed_hybrid_csds}, Fig.~\ref{fig:ed_ectcar}a, Fig.~\ref{fig:ed_transport}a--c and e--h  were measured using the arbitrary waveform generator. Finally, we plot the response of the resonators in 2D measurements by projecting it onto the principal component axis.
Further details of the fridge lines and circuitry can be found in Fig.~\ref{fig:ed_measurementsetup}.
The measurements reported in Fig.~\ref{fig:ed_pmmqubit} were performed on a similar setup.

\subsection{Device tune-up}\label{Methods_tuneup}
For tuning up the device, we follow a procedure that is described in ref~\cite{zatelli2024robust}. The device is controlled using 11 bottom gates that define electrostatic confinement, while a parallel magnetic field $B = 150$\,mT ensures spin polarization of the quantum dots~\cite{dvir2023pmm}. 
First, we perform transport measurements to confirm the presence of Andreev bound states in the hybrid segment, using $V_\mathrm{LT}$ and $V_\mathrm{RT}$ as tunnel barriers. Next, we form a pair of quantum dots QDL and QDR adjacent to the hybrid by confining quantum levels around $\VQDR$ and $\VQDL$ with the adjacent gates. We proceed to increase their coupling to the hybrid by lowering the tunnel barrier between them using $V_\mathrm{LT}$ and $V_\mathrm{RT}$, such that the quantum dot levels hybridize with the Andreev bound states. 
We form the additional quantum dot (CS, controlled with $\VCS$) on the right side that acts as a single-lead charge sensor for QDR~\cite{persson2010excess,bruhat2016cavity,vandrielPRXQuantum.5.020301}. We deplete the nanowire in between QDR and CS by setting $V_\mathrm{cut,CS} = -1.2$\,V while tracking the QDR and CS resonances. As such, CS is capacitively coupled to QDR without tunnel coupling while residual capacitive coupling to QDL is negligible due to screening by the superconducting segment~\cite{vandrielPRXQuantum.5.020301}.
We subsequently take charge-stability diagrams of the QDs with $\VQDR$ and $\VQDL$ while adjusting $\VH$, to observe hybridization between the quantum dot levels. The coupling between QDL and QDR is mediated by the Andreev bound states in the hybrid, with crossed-Andreev (CAR) and elastic co-tunneling (ECT) amplitudes that depend on the hybrid potential and can be balanced by tuning $\VH$~\cite{dvir2023pmm}. We use this to identify PMM sweet spots in the transport regime. 
Finally, we deplete the left part of the nanowire using $\Vcut$ to isolate the chain from the left lead, while keeping track of QDL's resonance. The device in Fig.~\ref{fig:ed_pmmqubit} was tuned up following a similar procedure.

\subsection{Time traces measurement and analysis}\label{Methods_timetrace}
The time traces presented in this text are measured by continuously recording the complex demodulated reflected signal for $\SI{10}{s}$, except for Fig.~\ref{fig:ed_weakcoupling} and Fig.~\ref{fig:ed_pmmqubit}, where we recorded $\SI{5}{s}$ and $\SI{2}{s}$ time traces, respectively. The measurement bandwidth is $\SI{100}{\kilo Hz}$ for Fig.~\ref{fig:3}c,d and Fig.~\ref{fig:ed_power_extra_analysis}c--g, $\SI{50}{\kilo Hz}$ for Fig.~\ref{fig:ed_power_extra_analysis}a,b, $\SI{20}{\kilo Hz}$ for Fig.~\ref{fig:3}e, Fig.~\ref{fig:4}, Fig.~\ref{fig:ed_ectcar}, Fig.~\ref{fig:ed_chargesensor}e--i, and Fig.~\ref{fig:ed_sweetspot2}, $\sim \SI{17}{\kilo Hz}$ for Fig.~\ref{fig:ed_pmmqubit}, and $\SI{10}{\kilo Hz}$ for Fig.~\ref{fig:ed_weakcoupling}.
We analyze each complex time trace by further averaging it in time bins of $\SI{150}{\micro s}$ to improve the SNR, unless otherwise specified. Afterwards, we project it onto the principal component axis. To unambiguously determine the principal component vector, we assume its real component to be positive.
The resulting one-dimensional time trace is then split in 10 subtraces. We fit each subtrace with a two-state hidden Markov model with Gaussian emissions using the Python package \textit{hmmlearn}. The fitted Markov model assigns the most likely sequence of states to the data and estimates the transition probabilities between them. The two states are then mapped to the physical ``even" and ``odd" state as described in Methods section \ref{Methods_state_assignment}.
From the transition probabilities $p_\mathrm{e\rightarrow o}, p_\mathrm{o\rightarrow e}$ we calculate the corresponding transition rates as $\Gamma_\mathrm{e \rightarrow o,o\rightarrow e}=-\log(1-p_\mathrm{o\rightarrow e,e\rightarrow o})/\tau_\mathrm{bin}$. These can be used to estimate other relevant quantities such as the lifetimes $\tau_\mathrm{e} = 1/\Gamma_\mathrm{e \rightarrow o}$, $\tau_\mathrm{o} = 1/\Gamma_\mathrm{o \rightarrow e}$, $\tau_\mathrm{avg}=2/(\Gamma_\mathrm{e \rightarrow o}+\Gamma_\mathrm{o \rightarrow e})$ and the parity polarization $(\tau_\mathrm{e}-\tau_\mathrm{o})/(\tau_\mathrm{e}+\tau_\mathrm{o})$. Additionally, the average and standard deviation of the reflected signal can be used to estimate the center and width of the Gaussians shown in Fig.~\ref{fig:3}d.
Finally, we average the results of each subtrace to estimate the corresponding quantities for the whole time trace. We consider the standard deviation of the mean as their uncertainty.
Importantly, this analysis relies on the assumption that the switching processes are Poissonian~\cite{hays2018direct}. To verify this, in Fig.~\ref{fig:ed_power_extra_analysis}e we plot the distribution of the dwell times for the time trace discussed in Fig.~\ref{fig:3}c,d. The fit is in good agreement with an exponential distribution, as expected from a Poissonian process.
Additionally, we note that the choice of integration time can affect the estimation of the parity lifetime, as shown in Fig.~\ref{fig:ed_power_extra_analysis}f. Such a systematic error can occur if the integration time is not sufficiently smaller than the typical dwell time~\cite{naaman2006poisson}. Using a shorter integration time would be detrimental for the SNR and state assignment, as shown in Fig.~\ref{fig:ed_power_extra_analysis}c,d.

To further support our analysis, we extract the average switching time of the time trace in Fig.~\ref{fig:3}c,d using a different procedure that does not require averaging.
We divide the unaveraged time trace in 10 subtraces and calculate the power spectral density of each subtrace, using Welch's method as implemented in \textit{scipy}. We fit each power spectral density using a Lorentzian with an additional background:
\begin{equation}
    \mathrm{PSD}(f) = 4A\frac{\Gamma_\mathrm{avg}}{(2\Gamma_\mathrm{avg})^2+(2\pi f)^2}+B
\end{equation}
In Fig.~\ref{fig:ed_power_extra_analysis}g we plot the average power spectral density of the 10 subtraces. We obtain an average switching time of $\tau_\mathrm{avg}=1/\Gamma_\mathrm{avg}=\SI{1.51(0.07)}{\milli s}$. This deviates by less than $\SI{0.5}{\milli s}$ from the estimation discussed in the main text. Although the finite integration time gives rise to a systematic error in the lifetime estimation, it does not affect the conclusions of the analysis.

To provide additional corroboration, we analyze the time trace of Fig.~\ref{fig:3}c using the methods and a minimally modified version of the code of Ref.~\cite{aghaee2025interferometric}. The results (Fig.~\ref{fig:ed_power_extra_analysis}h,i) are consistent with the hidden Markov model analysis.

Finally, we argue that the imperfect correlation coefficients presented in Fig.~\ref{fig:4} can be attributed to suboptimal SNR. To simplify the analysis, we assume that the parity polarization does not significantly deviate from 0. 
Let $X$ and $Y$ be two telegraph processes whose values can be $\pm 1$. If they are perfectly correlated, their correlation is $\rho_{XY}=\langle XY \rangle = 1$. In the presence of assignment errors, the observed signals are $X_{obs}$ and $Y_{obs}$. If the assignment errors are uncorrelated and their probability is $p_X$ and $p_Y$, the correlation is \mbox{$\rho_{X_{obs}Y_{obs}} = \langle X_{obs} Y_{obs} \rangle = 1 - p_X-p_Y+2p_X p_Y$}.
For the positively correlated trace in Fig.~\ref{fig:4}, we extracted $\mathrm{SNR_M}=1.27$ and $\mathrm{SNR_R}=0.97$. The corresponding assignment error is $p_{M}=\left[1-\mathrm{erf}(\mathrm{SNR_{M}/\sqrt{2}})\right]/2\approx 0.10$ and $p_{R}\approx 0.17$, resulting in an expected correlation of $\rho_\mathrm{MR} \approx 0.75$, close to the measured value.

To analyze the time trace measured in a qubit-geometry device (Fig.~\ref{fig:ed_pmmqubit}g–j), we follow a similar procedure. However, because of the additional complexity of assigning multiple states, we employ a three-state Gaussian mixture model instead of a hidden Markov model. To improve the signal-to-noise ratio (SNR), we apply a Gaussian filter ($\SI{100}{\micro s}$ width) rather than a box filter. The lifetimes of the three states and their uncertainties are obtained from the mean dwell times and the standard error of the mean. Transition rates between states $i$ and $j$ are calculated as $\Gamma_{i\rightarrow j} = N_{ij}/T_i$, where $N_{ij}$ is the number of transitions from state $i$ to $j$, and $T_i$ is the total time spent in state $i$. The uncertainty on the transition rates is estimated as $\sqrt{N_{ij}}/T_i$. If no transitions are observed, we report an upper bound based on the Garwood interval at the $95\%$ confidence level.

\subsection{State assignment}\label{Methods_state_assignment}

The two states of the Markov model are mapped to the physical ``even" and ``odd" state using the following procedures.

For the charge sensing measurements we assume that the state is even at the left of the charge-stability diagram, where the measured dot is empty. We then use the same mapping for the full charge-stability diagram.

For the quantum capacitance measurements, we assign the states such that the top-left and bottom-right corners of the charge-stability diagram are odd. This mapping is then used for the whole central region where the SNR is finite. In the regions with finite SNR that are disconnected from the central region (bottom-left and top-right of the charge-stability diagram), it is the odd state that gives rise to finite quantum capacitance, as further discussed in Methods section~\ref{Methods:Theory}. Therefore, we invert the mapping for these regions.
Practically, we identify the boundaries of the central region considering the contours $\mathrm{SNR_M} = 0.5$. Beyond these contours, the inverted mapping is used. We include a padding of 3 pixels to account for small inaccuracies in the definition of the boundaries. 
Although the threshold choice is somewhat arbitrary, we find that if $\mathrm{SNR_M} \lesssim  0.5$ the parity polarization varies significantly from pixel to pixel. This suggests that the analysis is not reliable if $\mathrm{SNR_M} \lesssim  0.5$. The same holds for the charge sensing measurements.

\subsection{Model}\label{Methods:Theory}
\setcounter{figure}{0}
\renewcommand{\thefigure}{S\arabic{figure}}
\renewcommand{\theHfigure}{S\arabic{figure}}
In this section we introduce a phenomenological model of the minimal Kitaev chain and quantum state dynamics, which we compare to experimental findings. First, we assume that the main contribution of CAR and ECT stems from an ABS state residing in the Al-covered part of the semiconductor, which we model as a proximitized resonant level. For simplicity, we neglect the rest of the superconducting continuum similar to Refs.~\cite{liu2022tunable, zatelli2024robust}. This results in the following QD-ABS-QD Hamiltonian,
\begin{align}
&H = H_{QD} + H_{ABS} + H_{T}, \label{Seq:1} \\
&H_{QD} = -\mu_{\mathrm{LD}}d_{L}^\dagger d_L -\mu_{\mathrm{RD}}d_R^\dagger d_R, \\
&H_{ABS} = -\mu_C \sum_\sigma d^\dagger_{C\sigma}d_{C\sigma}+\Delta_P\left(d^\dagger_{C\uparrow}d^\dagger_{C\downarrow}+d_{C\downarrow}d_{C\uparrow}\right)  \\
&H_T = \sum_{j=L/R} \left[t_{j\uparrow}d^\dagger_jd_{C\uparrow}+t_{j\downarrow} d^\dagger_jd_{C\downarrow} + \text{h.c.}\right].
\end{align}
Here, $d_L$ and $d_R$ are annihilation operators for the  spin-polarized left/right QDs, and $d_{C\sigma}$ for the central spin-full resonant level which is proximitized with superconducting pairing $\Delta_P$. A small Zeeman splitting is expected for the resonant level, as it is partly screened by the superconductor, which we for simplicity neglect in this analysis~\cite{liu2022tunable}.  Chemical potentials are given by $\mu_\mathrm{LD}$, $\mu_\mathrm{RD}$, and $\mu_C$. Now, by performing a Bogoliubov transformation, $d_{C\uparrow} = u\beta_\uparrow-v\beta^\dagger_{\downarrow}$ and $d_{C\downarrow}=u\beta_{\downarrow}+v\beta^\dagger_{\uparrow}$, we obtain,
\begin{align}
&H_{ABS} = E_\mathrm{ABS}\sum_\sigma\beta^\dagger_\sigma \beta_\sigma, \\
&H_T =\sum_{j=L/R}\left[t_{j\uparrow} d_j^\dagger(u\beta_\uparrow-v\beta^\dagger_\downarrow) + t_{j\downarrow} d_j^\dagger(u\beta_\downarrow+v\beta^\dagger_\uparrow) + \text{h.c.}\right],
\end{align}
where $E_\mathrm{ABS}=\sqrt{\mu_C^2+\Delta_P}$ is the effective gap of the ABS, and $u=\frac{1}{\sqrt{2}}\frac{-\Delta_P/E_\mathrm{ABS}}{\sqrt{1-\mu_C/E_\mathrm{ABS}}}$ and $v=\frac{1}{\sqrt{2}}\sqrt{1-\mu_C/E_\mathrm{ABS}}$ are the coherence factors. The tunneling, $t_{j\sigma}$, contains spin-dependence due to spin-orbit interaction. Assuming that both left and right QDs are spin-polarized in the down direction, $d_L=d_{L\downarrow}$, then $t_{L\downarrow}$ denotes normal tunneling and $t_{L\uparrow}$ spin-flipping tunneling. Next, to obtain the typical minimal Kitaev chain Hamiltonian we use quasi-degenerate perturbation theory to trace out the ABS levels and find
\begin{align}
&H_{0} = H_{QD} + \Delta d^\dagger_Ld^\dagger_R+td^\dagger_L d_R + \text{h.c.}, \label{Seq:HPMMapprox} \\
&\Delta = -\frac{2uv}{E_\mathrm{ABS}}\left(t_{L\uparrow}t_{R\downarrow}-t_{L\downarrow}t_{R\uparrow}\right), \\
&t = \frac{u^2-v^2}{E_\mathrm{ABS}}\left(t_{L\uparrow}t_{R\uparrow}+t_{L\downarrow}t_{R\downarrow}\right),
\end{align}
which is valid for $|E_\mathrm{ABS}\pm\mu_j| \ll |t_{j\sigma}|$ for $j=L/R$.
To obtain the MZM sweet spot from this model, we require $|t| = |\Delta|$ which is fulfilled by $2uv=\pm \left(t_{L\uparrow}t_{R\uparrow}+t_{L\downarrow}t_{R\downarrow}\right)/t_Lt_R$ with $t_j = \sqrt{t_{j\uparrow}^2+t^2_{j\downarrow}}$. This yields a gap to the excited states at the sweet spot
\begin{equation}
2|\Delta| = \frac{2t_L t_R}{E_\mathrm{ABS}}\left|2uv\left(u^2-v^2\right)\right|=\frac{t_L t_R}{E_\mathrm{ABS}}|\sin4\theta|,
\end{equation}
at chemical potentials $\mu_\mathrm{LD} = t_L^2\frac{u^2-v^2}{E_\mathrm{ABS}}$ and $\mu_\mathrm{RD}= t_R^2\frac{u^2-v^2}{E_\mathrm{ABS}}$. In the last expression, the angle $\theta$ connects to the tunnel couplings in the following way; $t_{L\uparrow} =t_L \cos\theta$, $t_{R\uparrow}=t_R\cos\theta$, $t_{L\downarrow}=t_L\sin\theta$, $t_{R\downarrow}=-t_R\sin\theta$, and the gap is maximized if $\theta =\pi/8$. In general however, the ratios $t_{j\uparrow}/t_{j\downarrow}$, and therefore $\theta$, are geometrically set in experiment and not gate-tuneable. The primary strategy is therefore to tune $\mu_C$ via electrostatic gating, which tunes $u$ and $v$, until a sweet spot is observed~\cite{dvir2023pmm}. In general, both the full Hamiltonian (Eq.~\ref{Seq:1}) and the minimal Kitaev Hamiltonian, (Eq.~\ref{Seq:HPMMapprox}) can be separated into even and odd parity sectors, with $E_{en}$ and $E_{on}$ denoting the $n$th excitation in the respective parity sector. For the Kitaev model, the lowest energy states in each parity are given by $E_{e0} = -\delta -\sqrt{\delta^2+\Delta}\equiv E_\mathrm{e}^-$ and $E_{o0}= \textcolor{red}{-\delta}-\sqrt{\varepsilon^2+t^2}\equiv E_\mathrm{o}^-$, where $\delta=(\mu_\mathrm{LD}+\mu_\mathrm{RD})/2$ and $\varepsilon=(\mu_\mathrm{LD}-\mu_\mathrm{RD})/2$, with excited states at $E_{e1}=-2\delta-E_{e0}$ and $E_{o1}=-2\delta-E_{o0}$, which are used in the main text. The theory plots shown in the paper are made by diagonalizing Eq.~\ref{Seq:1}, to ensure that details of the ABS beyond perturbation are captured, but we exclude the excited ABS singlet state of energy $2E_\mathrm{ABS}$ to stay consistent with us neglecting the BCS gap of comparable energy. The perturbative model, Eq.~\ref{Seq:HPMMapprox}, is used to derive minimal equations and gain intuition from the results. 

\subsubsection{Quantum capacitance and charge sensing}\label{Methods:TheoryQCCS}
In this subsection, we focus on the primary measures used in the experiment, namely the signals from quantum capacitance measured on the superconducting lead, and charge sensing of the right quantum dot. To derive the quantum capacitance arising from a small oscillating voltage $V_S$ on the superconductor, we ask how the energy of state $n$ changes with small fluctuations $\delta V_S$. Expanding to second order, we get
\begin{align}
E_n(V_{S}+\delta V_S)&\approx E_n(V_S) +E_n'(V_S)\delta V_S+\frac{1}{2}E_n''(V_S)(\delta V_S)^2 \\
&= \frac{1}{2}E_n''(V_S)\left(\delta V_S +\frac{E_n'(V_S)}{E_n''(V_S)}\right)^2 + E_n(V_S)-\frac{1}{2}\frac{E_n'(V_S)^2}{E_n''(V_S)},
\end{align}
where in the second line we recognize $C_n=E_n''(V_S)$ as a capacitance, as it matches the quadratic response $E_C = 1/2 C \delta V^2$ of a regular capacitor~\cite{Secchi2023Apr}. We note that this yields the capacitive response of a given state $n$ in a closed system, but neglects responses arising from opening the system to a bath, such as decoherence or change of state~\cite{peri2024unified}. We consider this appropriate as poisoning is slow compared to quantum capacitance measurements. Next, to evaluate $E_n''$, we consider the effects of $\delta V_S$ on chemical potentials, 
\begin{align}
&\mu_j(V_S+\delta V_S) \approx \mu_j(V_S) +(\alpha_{jS}-1)e\delta V_S,
\hspace{0.2cm} \Rightarrow \hspace{0.2cm} \frac{\partial \mu_j}{\partial V_S} = e\left(\alpha_{jS} -1\right) 
\end{align}
where we subtracted $e\delta V_S$ since we measure chemical potentials relative to the superconductor. Here, $\alpha_{jS}$ specifies the lever arm of the superconductor. Physically, $\alpha_{jS}=0$ relates to the chemical potentials being completely set by bottom gates, while $\alpha_{jS}=1$ means that the chemical potentials always matches the superconductor, such that $\delta V_S$ impose no relative change to $\mu_j$. For the ABS we set $\alpha_{CS}=1$ as we assume a strong capacitive coupling to the superconducting lead, thus fixing its chemical potential to the superconductor. For the QDs, $\alpha_{LS}$, $\alpha_{RS}\neq 1$, as the bottom gates primarily set their chemical potential. As such, we obtain
\begin{align}
E_n'' \approx &\frac{e^2}{4}(2 - \alpha_{LS}-\alpha_{RS})^2\frac{\partial^2E_n}{\partial\delta^2} + \frac{e^2}{4}(\alpha_{LS}-\alpha_{RS})^2\frac{\partial^2E_n}{\partial\varepsilon^2} \nonumber \\+ &\frac{e^2}{2}\frac{\partial^2E_n}{\partial\delta\partial\varepsilon}\left(2-\alpha_{LS}-\alpha_{RS}\right)\left(\alpha_{LS}-\alpha_{RS}\right). \label{seq:FullCapa}
\end{align}
Finally, assuming that lever arms are fairly symmetrical ($\alpha_{LS} \approx \alpha_{RS}$) and defining $\alpha = 2-\alpha_{LS}-\alpha_{RS}$, we use the Hellmann-Feynman theorem to rewrite
\begin{equation}
E''_n(V_S) \approx \frac{\alpha^2e^2}{4}\frac{\partial^2E_n}{\partial\delta^2} =\frac{\alpha^2e^2}{4}\frac{\partial}{\partial\delta}\bra{n}\frac{\partial H}{\partial\delta}\ket{n}.
\end{equation}
Here, $\partial H/\partial \delta$ is simply the total charge of both QDs (see Eq.~\ref{Seq:1}), and we arrive at the quantum capacitance of state $n$
\begin{equation}
C_{n} \approx \frac{\alpha^2e^2}{4}\frac{\partial^2E_n}{\partial\delta^2}= -\frac{\alpha^2e^2}{2}\frac{\partial}{\partial\delta}
\bra{n}d^\dagger_Ld_L+d^\dagger_Rd_R\ket{n}. \label{Seq:FinalCapa}
\end{equation}
Using this equation for the minimal Hamiltonian (Eq.~\ref{Seq:HPMMapprox}), the quantum capacitance of the odd states is zero, as both are superpositions of $\ket{01}$ and $\ket{10}$ with a constant total charge of $1$. The even ground state yields a signal
\begin{equation}
C_{e0} = -\alpha^2e^2\frac{1}{4}\frac{\Delta^2}{\left(\delta^2+\Delta^2\right)^\frac{3}{2}}, \label{Seq:PMMCapa}
\end{equation}
corresponding to a minimum around $\delta\approx0$ of width $\Delta$. We note that $\alpha_{LS}\neq \alpha_{RS}$ would also yield a signal for the odd state,
\begin{equation}
C_{o0} = -\left(\alpha_{LS}-\alpha_{RS}\right)^2e^2\frac{1}{4}\frac{t^2}{\left(\varepsilon^2+t^2\right)^\frac{3}{2}}
\end{equation}
along the line $\varepsilon\approx 0$ of width $t$ arising from the second term in Eq.~\ref{seq:FullCapa}. That this signal is not observed in experiment supports our assumption of $\alpha_{LS}\approx\alpha_{RS}$. In comparison, if the resonator was coupled to only one QD, through e.g. the left QD bottom gate, we would expect $\alpha_{RS}\approx\alpha_{CS}\approx1$ and $\alpha_{LS}\approx0$ as only $\mu_\mathrm{LD}$ varies relative to the superconductor. This would result in no parity distinguishing signal at the sweet spot, $|\Delta|=|t|$, since $C_{e0}=C_{o0}$, highlighting that read-out requires coupling of the Majoranas. In Fig.~\ref{fig:ed_theory}a, we plot the numerically obtained quantum capacitance signal, assuming $\alpha_{LS}=\alpha_{RS}$, in terms of the difference $\delta C_\mathrm{Q} = C_{o0}-C_{e0}$, which is appropriate as both $C_{e0}$ and $C_{o0}$ are strictly negative for Eq.~\ref{Seq:1} due to QD charge monotonically increasing with $\delta$ in Eq.~\ref{Seq:FinalCapa}. This quantity thus shows which state yields the dominant quantum capacitance signal, while areas of $\delta C_\mathrm{Q}=0$ indicate where the states are indistinguishable, i.e. yielding signals of equal amplitude. The positive signal along the anti-diagonal $\mu_\mathrm{LD}\approx-\mu_\mathrm{RD}$ stems from Eq.~\ref{Seq:PMMCapa}, while the negative corners stem from the odd state hybridizing with the Andreev bound state when $E_\mathrm{ABS}\approx E_{o0}-E_{e0}$, which goes beyond the minimal Kitaev chain Hamiltonian. To quantify how accurate the perturbative PMM model is at the sweet spot, we compare the numerically obtained signal strength, $C_{e0}-C_{o0}$, with the perturbative prediction of Eq.~\ref{Seq:PMMCapa}, $C_{e0,per}-C_{o0,per}=-\alpha^2e^2/4\sqrt{\Delta}$. For parameters $t_L=t_R=0.5E_\mathrm{ABS}$, $\theta=\pi/8$, and $\mu_C=0.75\Delta_P$ yielding a gap of $|\Delta|\approx 0.125E_\mathrm{ABS}$ we find $\left(C_{e0,per}-C_{o0,per}\right)/\left(C_{e0}-C_{o0}\right)\approx1.1$. This indicates that higher order terms reduce the signal slightly, on the scale of $10\%$.

For charge sensing, it is the charge of the right quantum dot that is measured, which for state $n$ is given by
\begin{equation}
Q_{R,n} = \bra{n}d^\dagger_Rd_R\ket{n}.
\end{equation}
For the minimal Kitaev Hamiltonian, Eq.~\ref{Seq:HPMMapprox}, this yields the signals
\begin{align}
Q_{R, e0} = \frac{\Delta^2}{2\sqrt{\delta^2+\Delta^2}}\frac{1}{\sqrt{\delta^2+\Delta^2}+\delta}, \hspace{0.4cm} Q_{R,o0} =\frac{t^2}{2\sqrt{\varepsilon^2+t^2}}\frac{1}{\sqrt{\varepsilon^2+t^2}+\varepsilon},
\end{align}
These quantities are strictly positive and describe an anti-diagonal and diagonal line, respectively, separating plateaus of $Q_R = 1$ from $Q_R =0$. Subtracting the lines ($\delta Q_\mathrm{R}=Q_{R,o0}-Q_{R,e0}$) yields two characteristic triangular shapes as shown in Fig.~\ref{fig:ed_theory}b. To identify the line of indistinguishability via charge sensing, we look for the condition
\begin{equation}
Q_{R,e0}= Q_{R,o0} \hspace{0.2cm} \Rightarrow \hspace{0.2cm} \Delta\left(\sqrt{\varepsilon^2+t^2}+\varepsilon\right)=t\left(\sqrt{\delta^2+\Delta^2}+\delta\right), \label{Seq:linecheck}
\end{equation}
which is fulfilled by $\mu_\mathrm{LD}=\frac{\Delta-t}{\Delta+t}\mu_\mathrm{RD}$. This can be verified by using the corresponding relations $\delta=\Delta\mu_\mathrm{RD}/(\Delta+t)$ and $\varepsilon=t\mu_\mathrm{RD}/(\Delta+t)$ in Eq.~\ref{Seq:linecheck}. This equation yields an indicator for the sweet spot, as the criterion of $t=\Delta$ yields a flat line separating the triangles, while other choices yields a titled line (Fig.~\ref{fig:ed_theory}c-e). This slope also correlates to the width of the $\delta Q_\mathrm{R}$ transitions, where the anti-diagonal width is given by $\Delta$ and the diagonal by $t$, meaning that the line tilts towards the diagonal of lowest width. Focusing now on the sweet spot ($\Delta=t$), we expand $\delta Q_\mathrm{R}$ around $\mu_\mathrm{LD}/\Delta$ for $\mu_{\mathrm{RD}}=0$, yielding
\begin{equation}
\delta Q_\mathrm{R}\approx\frac{-16\Delta\mu_{\mathrm{LD}}}{16\Delta^2+\mu_{\mathrm{LD}}^2},
\end{equation}
which corresponds to a Lorentzian with a full width at half maximum of $2\Delta$. The width of the white indistinguishability line separating the tips of the two triangles in Fig.~\ref{fig:ed_theory}c can thus be used to estimate the size of the gap at the sweet spot.

\subsubsection{Occupation}\label{Methods:TheoryOccupation}
Above, we investigated the minimal Hamiltonian and the signals arising from the lowest-energy parity states. Next, we wish to tackle the occupation of the even and odd parity states which goes beyond the closed-system Hamiltonian in Eq.~\ref{Seq:1}, and provide a minimal description for the observed parity polarizations and parity lifetimes. First, changing parity requires the addition or removal of a quasiparticle from the QD-ABS-QD system, which we assume to be the slowest rate in the system providing the millisecond parity lifetimes. Accordingly, we assume that the intra-parity relaxation rate is much faster~\cite{petta2004manipulation}, such that the system reaches the lowest-energy state in a given parity sector ($E_{e0}$ for even and $E_{o0}$ for odd) between parity switching events. We thus concentrate solely on the parity-changing rates which we assume to follow Fermi rates
\begin{align}
\Gamma_{e\rightarrow o} &= \Gamma_p n_F(E_{e0}-E_{o0},T_p) + \Gamma_p n_F(E_{e0}-E_{o1},T_p), \label{Seq:Rates} \\
\Gamma_{o\rightarrow e}&= \Gamma_pn_F(E_{o0}-E_{e0},T_p)+\Gamma_pn_F(E_{o0}-E_{e1},T_p).
\end{align}
Here, $\Gamma_p$ denotes the rate of poisoning quasiparticles that change the systems parity, $n_F(E,T)=1/\left(1+\exp E/ k_BT_p\right)$ the fermi distribution, and $T_p$ a phenomenological temperature added to capture smoothness of the parity polarization maps. This modeling is motivated by three observational facts: First, the parity polarizes towards the lowest-energy parity (specifically, towards even parity when $E_{e0}<E_{o0}$ and odd parity otherwise), as seen in Fig.~\ref{fig:3} of the main text. Second, the reflected signals $\SM$ and $\SR$ shows only signal from two states, meaning higher energy states than $E_{e0}$ and $E_{o0}$ are only negligibly occupied. Third, parity lifetimes decrease towards the central degeneracy point where the gap is lowest, as seen in Fig.~\ref{fig:ed_ectcar}. This indicates that the excited states $E_{e1}$ and $E_{o1}$ are involved in poisoning processes. 

From this model, the probabilities of being in the even and odd state are given by $P_{e} = \Gamma_{o\rightarrow e}/(\Gamma_{e\rightarrow o} + \Gamma_{o\rightarrow e})$ and $P_{o}=\Gamma_{e\rightarrow o}/(\Gamma_{e\rightarrow o} + \Gamma_{o\rightarrow e})$, with polarization given by $P_\mathrm{M}=P_e-P_o$ which is plotted in Fig.~\ref{fig:ed_theory}f--h for different values of $t$ and $\Delta$. The opening and closing of the cross mimics that of spectroscopy and corresponds to experiments shown in Fig.~\ref{fig:ed_ectcar}. In total, these equations mimic the system being coupled to a metallic bath with rate $\Gamma_p$ and temperature $T_p$. However, this does not result in a thermal equilibrium distribution at $T_p$ due to the assumption of fast intra-parity relaxation, such that only the lowest-energy parity states are populated. For $T_p\rightarrow 0$ the system polarizes completely in the lowest energy parity sector, with constant $\Gamma_{e\rightarrow o} + \Gamma_{o\rightarrow e} = \Gamma_p$. In the other limit of $k_BT_p\gg 2\Delta$, a peak in $\Gamma_{e\rightarrow o} + \Gamma_{o\rightarrow e} \approx 2\Gamma_p$ occurs at the sweet spot, corresponding to a dip in $\tau_{\mathrm{avg}}$ as shown in Fig.~\ref{fig:ed_theory}i. This is due to the added transitions, $\Gamma_\Delta = \Gamma_p n_F(E_{e0}-E_{o1},T)$ (and $e\leftrightarrow o)$, to the excited states, which at the vicinity of the sweet spot are minimal, $E_{e1}-E_{e0}\approx 2\Delta$ and $E_{o0}-E_{o1}\approx 2\Delta$, as sketched in Fig.~\ref{fig:ed_theory}j. Lastly, using this model we can also calculate the average quantum capacitance given by $\langle C\rangle = P_e C_{e0}+P_oC_{o0}$, which for small coupling is plotted in Fig.~\ref{fig:2}d of the main text.

\clearpage
\bibliography{sn-bibliography}% common bib file


%TC:endignore
\end{document}